\documentclass[sigconf,natbib]{acmart}

\PassOptionsToPackage{dvipsnames}{xcolor}

\usepackage{dsfont}
\usepackage{acronym}
\usepackage[noend]{algorithmic}
\usepackage{algorithm}
\usepackage{bbm}
\usepackage{beramono}
\usepackage{bm}
\usepackage{booktabs}
\usepackage[skip=0pt]{caption}
\usepackage{cleveref}
\usepackage[T1]{fontenc}
\usepackage{graphicx}
\usepackage{hyperref}
\usepackage{listings}
\usepackage{multirow}
\usepackage{natbib}
\usepackage{subcaption}
\usepackage{tabularx}
\usepackage[dvipsnames]{xcolor}
\usepackage{enumerate}
\usepackage[inline]{enumitem}
\usepackage{numprint}
\usepackage[normalem]{ulem}

\copyrightyear{2020}
\acmYear{2020}
\setcopyright{acmlicensed}\acmConference[SIGIR '20]{Proceedings of the 43rd International ACM SIGIR Conference on Research and Development in Information Retrieval}{July 25--30, 2020}{Virtual Event, China}
\acmBooktitle{Proceedings of the 43rd International ACM SIGIR Conference on Research and Development in Information Retrieval (SIGIR '20), July 25--30, 2020, Virtual Event, China}
\acmPrice{15.00}
\acmDOI{10.1145/3397271.3401102}
\acmISBN{978-1-4503-8016-4/20/07}

\definecolor{rj}{RGB}{0, 150, 0}
\definecolor{mdr}{RGB}{200, 0, 0}
\definecolor{ho}{RGB}{0, 50, 150}

\newcommand{\ranking}{R}
\newcommand{\displayranking}{\bar{R}}
\newcommand{\doc}{d}

\acrodef{IR}{Information Retrieval}
\acrodef{LTR}{Learning to Rank}
\acrodef{ARP}{Average Relevance Position}
\acrodef{DCG}{Discounted Cumulative Gain}
\acrodef{EM}{Expectation Maximization}

\theoremstyle{definition}

\theoremstyle{definition}

\allowdisplaybreaks

\author{Harrie Oosterhuis}
\affiliation{%
	\institution{University of Amsterdam}
	\city{Amsterdam}
	\country{The Netherlands}
}
\email{oosterhuis@uva.nl}
 
\author{Maarten de Rijke}
\orcid{0000-0002-1086-0202}
\affiliation{
 \institution{University of Amsterdam \& Ahold Delhaize}
 \city{Amsterdam}
 \country{The Netherlands}
}
\email{derijke@uva.nl}

\acmSubmissionID{66}

\title{Unbiased Learning to Rank in Compact Presentation Layouts}
\title{Unbiased Learning to Rank under Item Selection Bias}
\title{Stochastic Methods for Unbiased Learning to Rank}
\title{Stochastic Unbiased Learning to Rank for Item Selection Bias}
\title{Stochastic Unbiased Learning to Rank for Top-$k$ Rankings}
\title{Unbiased Learning to Rank for Top-$k$ Rankings}
\title{Policy-Aware Unbiased Learning to Rank for Top-$k$ Rankings}

\allowdisplaybreaks

\begin{document}

\begin{abstract}
Counterfactual \ac{LTR} methods optimize ranking systems using logged user interactions that contain interaction biases.
Existing methods are only unbiased if users are presented with all relevant items in every ranking.
There is currently no existing counterfactual unbiased \ac{LTR} method for top-$k$ rankings.
We introduce a novel policy-aware counterfactual estimator for \ac{LTR} metrics that can account for the effect of a stochastic logging policy.
We prove that the policy-aware estimator is unbiased if every relevant item has a non-zero probability to appear in the top-$k$ ranking.
Our experimental results show that the performance of our estimator is not affected by the size of $k$: for any $k$, the policy-aware estimator reaches the same retrieval performance while learning from top-$k$ feedback as when learning from feedback on the full ranking.
Lastly, we introduce novel extensions of traditional \ac{LTR} methods to perform counterfactual \ac{LTR} and to optimize top-$k$ metrics.
Together, our contributions introduce the first policy-aware unbiased \ac{LTR} approach that learns from top-$k$ feedback and optimizes top-$k$ metrics.
As a result, counterfactual \ac{LTR} is now applicable to the very prevalent top-$k$ ranking setting in search and recommendation.
\end{abstract}

\maketitle

\acresetall

\section{Introduction}
\label{sec:intro}

\ac{LTR} optimizes ranking systems to provide high quality rankings.
Interest in \ac{LTR} from user interactions has greatly increased in recent years with the introduction of unbiased \ac{LTR} methods~\citep{joachims2017unbiased, wang2016learning}.
The potential for learning from logged user interactions is great:
user interactions provide valuable implicit feedback while also being cheap and relatively easy to acquire at scale~\citep{joachims2017accurately}.
However, interaction logs also contain large amounts of bias, which is the result of both user behavior and the ranker used during logging.
For instance, users are more likely to examine items at the top of rankings, consequently the display position of an item heavily affects the number of interactions it receives~\citep{wang2018position}.
This effect is called \emph{position bias} and it is very dominant when learning from interactions with rankings.
Naively ignoring it during learning can be detrimental to ranking performance, as the learning process is strongly impacted by what rankings were displayed during logging instead of \emph{true} user preferences.
The goal of unbiased \ac{LTR} methods is to optimize a ranker w.r.t. the \emph{true} user preferences, consequently, they have to account and correct for such forms of bias.

Previous work on unbiased \ac{LTR} has mainly focussed on accounting for position bias through counterfactual learning~\citep{joachims2017unbiased, wang2016learning, ai2018unbiased}.
The prevalent approach models the probability of a user examining an item in a displayed ranking.
This probability can be inferred from user interactions~\citep{joachims2017unbiased, wang2016learning, ai2018unbiased, wang2018position, agarwal2019estimating} and corrected for using \emph{inverse propensity scoring}.
As a result, these methods optimize a loss that in expectation is unaffected by the examination probabilities during logging, hence it is unbiased w.r.t.\ position bias.

This approach has been applied effectively in various ranking settings, including search for scientific articles~\cite{joachims2017unbiased}, email~\cite{wang2016learning} or other personal documents~\cite{wang2018position}.
However, a limitation of existing approaches is that in every logged ranking they require every relevant item to have a non-zero chance of being examined~\cite{carterette2018offline, joachims2017unbiased}.
In this paper, we focus on top-$k$ rankings where the number of displayed items is systematically limited.
These rankings can display at most $k$ items, making it practically unavoidable that relevant items are missing.
Consequently, existing counterfactual \ac{LTR} methods are not unbiased in these settings.
We recognize this problem as \emph{item selection bias} introduced by the selection of (only) $k$ items to display.
This is especially concerning since top-$k$ rankings are quite prevalent, e.g., in recommendation~\citep{cremonesi2010performance, hurley2011novelty}, mobile search~\citep{balke2002real, vlachou2011monitoring}, query autocompletion~\citep{cai-survey-2016,wang2016learning, wang2018position}, and digital assistants~\citep{shalyminov-2018-neural}.

Our main contribution is a novel policy-aware estimator for counterfactual \ac{LTR} that accounts for both a stochastic logging policy and the users' examination behavior.
Our policy-aware approach can be viewed as a generalization of the existing counterfactual \ac{LTR} framework~\citep{joachims2017unbiased, agarwal2019counterfactual}.
We prove that our policy-aware approach performs unbiased \ac{LTR} and evaluation while learning from top-$k$ feedback.
Our experimental results show that while our policy-aware estimator is unaffected by the choice of $k$, the existing policy-oblivious approach is strongly affected even under large values of $k$.
For instance, optimization with the policy-aware estimator on top-5 feedback reaches the same performance as when receiving feedback on all results.
Furthermore, because top-$k$ metrics are the only relevant metrics in top-$k$ rankings, we also propose extensions to traditional \ac{LTR} approaches that are proven to optimize top-$k$ metrics unbiasedly and introduce a pragmatic way to choose optimally between available loss functions.

Our work is based around two main contributions:
\begin{enumerate}[align=left, leftmargin=*,topsep=1.5pt]
	\item A novel estimator for unbiased \ac{LTR} from top-$k$ feedback.
	\item Unbiased losses that optimize bounds on top-$k$ \ac{LTR} metrics.
\end{enumerate}
To the best of our knowledge, our policy-aware estimator is the first estimator that is unbiased in top-$k$ ranking settings.

\section{Background}
\label{sec:background}

In this section we discuss supervised \ac{LTR} and the existing counterfactual \ac{LTR} framework~\citep{joachims2017unbiased}.

\subsection{Supervised \ac{LTR}}
\label{section:supervisedLTR}
The goal of \ac{LTR} is to optimize ranking systems w.r.t.\ specific ranking metrics.
Ranking metrics generally involve items $\doc$, their relevance $r$ w.r.t.\ a query $q$, and their position in the ranking  $\ranking$ produced by the system.
We will optimize the \emph{Empirical Risk}~\citep{vapnik2013nature} over the set of queries $Q$, with a loss $\Delta(\ranking_i \mid q_i, r)$ for a single query $q_i$:
\begin{align}
\mathcal{L} = \frac{1}{|Q|} \sum_{q_i \in Q} \Delta(\ranking_i \mid q_i, r).
\end{align}
For simplicity we assume that relevance is binary:
$r(q,\doc) \in \{0, 1\}$; for brevity we write: $r(q, \doc) = r(\doc)$.
Then, ranking metrics commonly take the form of a sum over items:
\begin{align}
\Delta(\ranking \mid q, r) =
\sum_{\doc \in \ranking} \lambda\left(\doc \mid \ranking \right) \cdot r(\doc),
\end{align}
where $\lambda$ can be chosen for a specific metric, e.g., for \ac{ARP} or \ac{DCG}:
\begin{align}
\lambda^\textit{ARP}(\doc \mid \ranking) &= \text{rank}(\doc \mid \ranking) \label{eq:ARP},\\
\lambda^\textit{DCG}(\doc \mid \ranking) &= -\log_2\big(1 + \text{rank}(\doc \mid \ranking)\big)^{-1}.  \label{eq:DCG}
\end{align}
In a so-called \emph{full-information} setting, where the relevance values $r$ are known, optimization can be done through traditional \ac{LTR} methods~\citep{wang2018lambdaloss, burges2010ranknet, joachims2002optimizing, liu2009learning}.

\subsection{Counterfactual \ac{LTR}}
\label{section:counterfactualLTR}
Optimizing a ranking loss from the implicit feedback in interaction logs requires a different approach from supervised \ac{LTR}. %
We will assume that clicks are gathered using a logging policy $\pi$ with the probability of displaying ranking $\displayranking$ for query $q$ denoted as $\pi(\displayranking \mid q)$.
Let $o_i(\doc) \in \{0, 1\}$ indicate whether $\doc$ was examined by a user at interaction $i$ and $o_i(\doc) \sim P( o(\doc) \mid q_i, r, \displayranking_i)$.
Furthermore, we assume that users click on all relevant items they observe and nothing else: $c_i(\doc) = \mathds{1}[r(\doc) \land o_i(\doc)]$.
Our goal is to find an estimator $\hat{\Delta}$ that provides an unbiased estimate of the actual loss; for $N$ interactions this estimate is:
\begin{align}
\hat{\mathcal{L}} =
\frac{1}{N} \sum^N_{i=1} \hat{\Delta}(\ranking_i\mid q_i, \displayranking_i, \pi, c_i). \label{eq:highlevelloss}
\end{align}
We write $\ranking_i$ for the ranking produced by the system for which the loss is being computed, while $\displayranking_i$ is the ranking that was displayed when logging interaction $i$.
For brevity we will drop $i$ from our notation when only a single interaction is involved.
A naive estimator could simply consider every click to indicate relevance:
\begin{align}
\hat{\Delta}_\textit{naive}\left(\ranking \mid q, c \right)
= \sum_{\doc : c(\doc) = 1} \lambda\left(\doc \mid \ranking \right)
.
\label{eq:naiveestimator}
\end{align}
Taking the expectation over the displayed ranking and observance variables results in the following expected loss:
\begin{align}
\mathbb{E}&_{o,\displayranking}\left[ \hat{\Delta}_\textit{naive}\left(\ranking \mid  q, c\right) \right] 
\nonumber \\
& =
\mathbb{E}_{o,\displayranking}\left[ \sum_{\doc: c(\doc) = 1} \lambda( \doc \mid \ranking) \right]
=
\mathbb{E}_{o,\displayranking}\left[ \sum_{\doc \in \ranking} \lambda( \doc \mid \ranking) \cdot c(\doc) \right]
\nonumber \\
&=
\mathbb{E}_{o,\displayranking}\left[ \sum_{\doc \in \ranking} o(\doc) \cdot \lambda\left(\doc \mid \ranking\right) \cdot r(\doc)\right] 
 \\
&=
\mathbb{E}_{\displayranking}\left[ \sum_{\doc \in \ranking} P\left( o(\doc) = 1 \mid q, r, \displayranking \right) \cdot \lambda\left(\doc \mid \ranking \right) \cdot r(\doc)\right] \nonumber \\
& =
\sum_{\displayranking \in \pi(\cdot \mid q)} \pi(\displayranking \mid q) \cdot 
\sum_{\doc \in \ranking} P\left( o(\doc) = 1 \mid q, r, \displayranking\right) \cdot \lambda\left(\doc\mid \ranking \right) \cdot r(\doc).
\nonumber
\end{align}
Here, the effect of position bias is very clear; in expectation, items are weighted according to their probability of being examined.
Furthermore, it shows that examination probabilities are determined by both the logging policy $\pi$ and user behavior $P( o(\doc) \mid q, r, \displayranking )$.

In order to avoid the effect of position bias, \citeauthor{joachims2017unbiased}~\citep{joachims2017unbiased} introduced an inverse-propensity-scoring estimator in the same vain as previous work by \citeauthor{wang2016learning}~\citep{wang2016learning}.
The main idea behind this estimator is that if the examination probabilities are known, then they can be corrected for per click:
\begin{align}
\hat{\Delta}_{\textit{oblivious}}\left(\ranking \mid q, c, \displayranking \right)
&= \sum_{\doc :c(\doc) = 1} \frac{\lambda\left(\doc \mid \ranking \right)}{P\left( o(\doc) = 1 \mid q, r, \displayranking \right) }.
\label{eq:obliviousestimator}
\end{align}
In contrast to the naive estimator (Eq.~\ref{eq:naiveestimator}), this policy-oblivious estimator (Eq.~\ref{eq:obliviousestimator}) can provide an unbiased estimate of the loss:
\begin{align}
\begin{split}
\mathbb{E}_{o,\displayranking}\Big[ &\hat{\Delta}_\textit{oblivious}\left(\ranking \mid q, c, \displayranking \right) \Big] \\
&= \mathbb{E}_{o,\displayranking} \Bigg[  \sum_{\doc:c(\doc) = 1} \frac{\lambda\left(\doc\mid \ranking \right)}{P\left( o(\doc) = 1 \mid q, r, \displayranking \right) } \Bigg] \\
&= \sum_{\doc \in \ranking} \mathbb{E}_{o,\displayranking} \Bigg[ \frac{o(\doc) \cdot \lambda\left(\doc\mid \ranking \right) \cdot r(\doc)}{P\left( o(\doc) = 1 \mid q, r, \displayranking \right)} \Bigg] \\
&= \sum_{\doc \in \ranking} \mathbb{E}_{\displayranking} \Bigg[ \frac{P\left( o(\doc) = 1 \mid q, r, \displayranking \right) \cdot \lambda\left(\doc\mid \ranking \right) \cdot r(\doc)}{P\left( o(\doc) = 1 \mid q, r, \displayranking \right)} \Bigg] \\
&= \sum_{\doc \in \ranking} \lambda\left(\doc\mid \ranking \right) \cdot r(\doc) 
= \Delta(\ranking\mid q, r).
\end{split}
\end{align}
We note that the last step assumes $P\left( o(\doc) = 1 \mid q, r, \displayranking \right) > 0$, and that only relevant items $r(\doc) = 1$ contribute to the estimate~\citep{joachims2017unbiased}.
Therefore, this estimator is unbiased as long as the examination probabilities are positive for every relevant item:
\begin{align}
\forall \doc , \,
\forall \displayranking \in \pi(\cdot \mid q) \,
\left[
r(\doc) = 1 \to P\left( o(\doc) = 1 \mid q, r, \displayranking \right) > 0
\right]
. 
\label{eq:agnosticcond}
\end{align}
Intuitively, this condition exists because propensity weighting is applied to items clicked in the displayed ranking and items that cannot be observed can never receive clicks.
Thus, there are no clicks that can be weighted more heavily to adjust for the zero observance probability of an item. 

An advantageous property of the policy-oblivious estimator $\hat{\Delta}_{\textit{oblivious}}$ is that the logging policy $\pi$ does not have to be known.
That is, as long as Condition~\ref{eq:agnosticcond} is met, it works regardless of how interactions were logged.
Additionally, \citeauthor{joachims2017unbiased}~\citep{joachims2017unbiased} proved that it is still unbiased under click noise. %
Virtually all recent counterfactual \ac{LTR} methods use the policy-oblivious estimator for \ac{LTR} optimization~\citep{wang2016learning, joachims2017unbiased, agarwal2019addressing, ai2018unbiased, wang2018position, agarwal2019estimating}.

\section{Learning from top-$k$ feedback}
\label{sec:topkfeedback}

In this section we explain why the existing policy-oblivious counterfactual \ac{LTR} framework is not applicable to top-$k$ rankings.
Subsequently, we propose a novel solution through policy-aware propensity scoring that takes the logging policy into account.

\subsection{The Problem with Top-$k$ Feedback}
An advantage of the existing policy-oblivious estimator for counterfactual \ac{LTR} described in Section~\ref{section:counterfactualLTR} is that the logging policy does not need to be known, making its application easier.
However, the policy-oblivious estimator is only unbiased when Condition~\ref{eq:agnosticcond} is met: every relevant item has a non-zero probability of being observed in every ranking displayed during logging.

We recognize that in top-$k$ rankings, where only $k$ items can be displayed, relevant items may systematically \emph{lack} non-zero examination probabilities.
This happens because items outside the top-$k$ cannot be examined by the user:
\begin{align}
\forall \doc, \forall \displayranking \,
\left[
\text{rank}\big(\doc \mid \displayranking \big) > k \to P\big( o(\doc) = 1 \mid q, r, \displayranking \big) = 0
\right]
.
\end{align}
In most top-$k$ ranking settings it is very unlikely that Condition~\ref{eq:agnosticcond} is satisfied; If $k$ is very small, the number of relevant items is large, or if the logging policy $\pi$ is ineffective at retrieving relevant items, it is unlikely that all relevant items will be displayed in the top-$k$ positions.
Moreover, for a small value of $k$ the performance of the logging policy $\pi$ has to be near ideal for all relevant items to be displayed.
We call this effect \emph{item selection bias}, because in this setting the logging ranker makes a selection of which $k$ items to display, in addition to the order in which to display them (position bias).
The existing policy-oblivious estimator for counterfactual \ac{LTR} (as described in Section~\ref{section:counterfactualLTR}) cannot correct for item selection bias when it occurs, and can thus be affected by this bias when applied to top-$k$ rankings.

\subsection{Policy-Aware Propensity Scoring}
Item selection bias is inevitable in a single top-$k$ ranking, due to the limited number of items that can be displayed.
However, across multiple top-$k$ rankings more than $k$ items could be displayed if the displayed rankings differ enough.
Thus, a stochastic logging-policy could provide every item with a non-zero probability to appear in the top-$k$ ranking.
Then, the probability of examination can be calculated as an expectation over the displayed ranking:
\begin{align}
P\left(o(\doc) = 1 \mid q, r, \pi \right)
&= \mathbb{E}_{\displayranking}\left[P\big(o(\doc) = 1 \mid q, r, \displayranking \big) \right] 
\label{eq:expexam} \\
&= \sum_{\displayranking \in \pi(\cdot \mid q)} \pi\big(\displayranking \mid q\big) \cdot  P\big(o(\doc) = 1 \mid q, r, \displayranking \big).
\nonumber
\end{align}
This policy-dependent examination probability can be non-zero for all items, even if all items cannot be displayed in a single top-$k$ ranking.
Naturally, this leads to a \emph{policy-aware} estimator:
\begin{align}
\hat{\Delta}_{\textit{aware}}\left(\ranking \mid q, c, \pi \right)
&= \sum_{\doc :c(\doc) = 1} \frac{\lambda(\doc \mid \ranking )}{P\big( o(\doc) = 1 \mid q, r, \pi \big) }. \label{eq:policyaware}
\end{align}
By basing the propensity on the policy instead of the individual rankings, the policy-aware estimator can correct for zero observance probabilities in some displayed rankings by more heavily weighting clicks on other displayed rankings with non-zero observance probabilities.
Thus, if a click occurs on an item that the logging policy rarely displays in a top-$k$ ranking, this click may be weighted more heavily than a click on an item that is displayed in the top-$k$ very often.
In contrast, the policy-oblivious approach only corrects for the observation probability for the displayed ranking in which the click occurred, thus it does not correct for the fact that an item may be missing from the top-$k$ in other displayed rankings.

In expectation, the policy-aware estimator provides an unbiased estimate of the ranking loss:
\begin{align}
\mathbb{E}_{o,\displayranking}\hspace{-5mm}&\hspace{5mm}\Big[\hat{\Delta}_\textit{aware}\left(\ranking \mid q, c, \pi \right) \Big] \nonumber \\
&
= \mathbb{E}_{o,\displayranking} \Bigg[  \sum_{\doc:c(\doc) = 1} \frac{\lambda\big(\doc \mid \ranking\big)}{P\left( o(\doc) = 1 \mid q, r, \pi\right) } \Bigg] \nonumber \\
&
= \sum_{\doc \in \ranking} \mathbb{E}_{o,\displayranking} \Bigg[ \frac{o(\doc) \cdot \lambda\big(\doc \mid \ranking\big) \cdot r(\doc)}{\sum_{\displayranking' \in \pi(\cdot \mid q)} \pi\big(\displayranking' \mid q\big) \cdot  P\big(o(\doc) = 1 \mid q, r, \displayranking' \big)} \Bigg]\\
&
= \sum_{\doc \in \ranking} \mathbb{E}_{\displayranking} \Bigg[ \frac{P\big( o(\doc) = 1 \mid q, r, \displayranking \big) \cdot \lambda\big(\doc \mid \ranking\big) \cdot r(\doc)}{\sum_{\displayranking' \in \pi(\cdot \mid q)} \pi\big(\displayranking' \mid q\big) \cdot  P\big(o(\doc) = 1 \mid q, r, \displayranking' \big)} \Bigg]  \nonumber \\
&
= \sum_{\doc \in \ranking}  \frac{\sum_{\displayranking \in \pi(\cdot \mid q)} \pi\big(\displayranking \mid q\big) \cdot  P\big( o(\doc) = 1 \mid q, r, \displayranking \big) \cdot \lambda\big(\doc \mid \ranking\big) \cdot r(\doc)}{\sum_{\displayranking'\in \pi(\cdot \mid q)} \pi\big(\displayranking' \mid q\big) \cdot  P\big(o(\doc) = 1 \mid q, r, \displayranking' \big)} \nonumber \\
&
= \sum_{\doc \in \ranking} \lambda\big(\doc \mid \ranking\big) \cdot r(\doc)
= \Delta\big(\ranking\mid q, r\big).
\nonumber
\end{align}
In contrast to the policy-oblivious approach (Section~\ref{section:counterfactualLTR}), this proof is sound as long as every relevant item has a non-zero probability of being examined under the logging policy $\pi$: 
\begin{align}
\forall \doc \,
\left[
 r(\doc) = 1 \to \sum_{\displayranking \in \pi(\cdot \mid q)} \pi\big(\displayranking \,|\, q\big) \cdot  P\big(o(\doc) = 1 \,|\, q, r, \displayranking \big) > 0
 \right]
 . 
 \label{eq:awarecond}
\end{align}
It is easy to see that Condition~\ref{eq:agnosticcond} implies Condition~\ref{eq:awarecond}, in other words, for all settings where the policy-oblivious estimator (Eq.~\ref{eq:obliviousestimator}) is unbiased, the policy-aware estimator (Eq.~\ref{eq:policyaware}) is also unbiased.
Conversely, Condition~\ref{eq:awarecond} does not imply Condition~\ref{eq:agnosticcond}, thus there are cases where the policy-aware estimator is unbiased but the policy-oblivious estimator is not guaranteed to be.

To better understand for which policies Condition~\ref{eq:awarecond} is satisfied, we introduce a substitute Condition~\ref{eq:awarecond2}:
\begin{align}
\mbox{}
\hspace*{-2mm}
\forall \doc 
\Big[
r(\doc) = 1 \to  \exists \displayranking \left[\pi\big(\displayranking \mid q\big)  > 0  \land  P\big(o(\doc) = 1 \mid q, r, \displayranking \big) > 0 \right]
\!
\Big]
. 
\hspace*{-2mm}
\mbox{}
\label{eq:awarecond2}
\end{align}
Since Condition~\ref{eq:awarecond2} is equivalent to Condition~\ref{eq:awarecond}, we see that 
the policy-aware estimator is unbiased for any logging-policy that provides a non-zero probability for every relevant item to appear in a position with a non-zero examination probability.
Thus to satisfy Condition~\ref{eq:awarecond2} in a top-$k$ ranking setting, every relevant item requires a non-zero probability of being displayed in the top-$k$.

As long as Condition~\ref{eq:awarecond2} is met, a wide variety of policies can be chosen according to different criteria.
Moreover, the policy can be deterministic if $k$ is large enough to display every relevant item.
Similarly, the policy-oblivious estimator can be seen as a special case of the policy-aware estimator where the policy is deterministic (or assumed to be).
The big advantage of our policy-aware estimator is that it is applicable to a much larger number of settings than the existing policy-oblivious estimator, including those were feedback is only received on the top-$k$ ranked items.

\subsection{Illustrative Example}

To better understand the difference between the policy-oblivious and policy-aware estimators, we introduce an illustrative example that contrasts the two.
We consider a single query $q$ and a logging policy $\pi$ that chooses between two rankings to display: $\displayranking_1$ and $\displayranking_2$, with: $\pi(\displayranking_1 \mid q) > 0$; $\pi(\displayranking_2 \mid q) > 0$; and $\pi(\displayranking_1 \mid q) + \pi(\displayranking_2 \mid q) = 1$.
Then for a generic estimator we consider how it treats a single relevant item $\doc_n$ with $r(\doc_n) \not= 0$ using the expectation:
\begin{align}
&\mathbb{E}_{o, \displayranking}\Big[
\frac{c(\doc_n) \cdot \lambda\big(\doc_n \mid \ranking\big)}{ \rho\big( o(\doc_n) = 1 \mid q, \doc_n, \displayranking, \pi\big)}
\Big]
= \lambda\big(\doc_n \mid \ranking\big) \cdot r(\doc_n) \cdot \phantom{x} \\
&\bigg( \frac{\pi(\displayranking_1 | q) \cdot P\big( o(\doc_n) = 1 | q, r, \displayranking_1 \big)}{ \rho\big( o(\doc_n) = 1 \mid q, \doc_n, \displayranking_1, \pi \big)}
+ \frac{\pi(\displayranking_2 | q) \cdot P\big( o(\doc_n) = 1 | q, r, \displayranking_2 \big)}{ \rho\big( o(\doc_n) = 1 \mid q, \doc_n, \displayranking_2,  \pi \big)}
\bigg), \nonumber
\end{align}
where the propensity function $\rho$ can be chosen to match either the policy-oblivious (Eq.~\ref{eq:obliviousestimator}) or policy-aware (Eq.~\ref{eq:policyaware}) estimator.

First, we examine the situation where $\doc_n$ appears in the top-$k$ of both rankings $\displayranking_1$ and $\displayranking_2$, thus it has a positive observance probability in both cases: $P\big( o(\doc_n) = 1 \mid q, r, \displayranking_1 \big) > 0$ and $P\big( o(\doc_n) = 1 \mid q, r, \displayranking_2 \big) > 0$.
Here, the policy-oblivious estimator $\hat{\Delta}_{\textit{oblivious}}$~(Eq.~\ref{eq:obliviousestimator}) removes the effect of observation bias by adjusting for the observance probability per displayed ranking:
\begin{align}
\bigg(& \frac{\pi(\displayranking_1 | q) \cdot P\big( o(\doc_n) = 1 | q, r, \displayranking_1 \big)}{ P\big( o(\doc_n) = 1 \mid q, r, \displayranking_1 \big)}
+ \frac{\pi(\displayranking_2 | q) \cdot P\big( o(\doc_n) = 1 | q, r, \displayranking_2 \big)}{ P\big( o(\doc_n) = 1 \mid q, r, \displayranking_2 \big)}
\bigg)  \nonumber \\
 &\cdot \lambda\big(\doc_n \mid \ranking\big) \cdot r(\doc_n) = \lambda\big(\doc_n \mid \ranking\big) \cdot r(\doc_n).
\end{align}
The policy-aware estimator $\hat{\Delta}_{\textit{aware}}$~(Eq.~\ref{eq:policyaware}) also corrects for the examination bias, but because its propensity scores are based on the policy instead of the individual rankings~(Eq.~\ref{eq:expexam}), it uses the same score for both rankings:
\begin{align}
& \frac{\pi(\displayranking_1 \mid q) \cdot P\big( o(\doc_n) = 1 |\, q, r, \displayranking_1 \big) + \pi(\displayranking_2 \mid q) \cdot P\big( o(\doc_n) = 1 |\, q, r, \displayranking_2 \big)}{
\pi(\displayranking_1 \mid q) \cdot P\big( o(\doc_n) = 1 |\, q, r, \displayranking_1 \big) + \pi(\displayranking_2 \mid q) \cdot P\big( o(\doc_n) = 1 |\, q, r, \displayranking_2 \big)
}
  \nonumber \\
 &\cdot \lambda\big(\doc_n \mid \ranking\big) \cdot r(\doc_n) = \lambda\big(\doc_n \mid \ranking\big) \cdot r(\doc_n).
\end{align}
Then, we consider a different relevant item $\doc_m$ with $r(\doc_m) = r(\doc_n)$ that unlike the previous situation only appears in the top-$k$ of $\displayranking_1$.
Thus it only has a positive observance probability in $\displayranking_1$: $P\big( o(\doc_m) = 1 \mid q, r, \displayranking_1 \big) > 0$ and $P\big( o(\doc_m) = 1 \mid q, r, \displayranking_2 \big) = 0$.
Consequently, no clicks will ever be received in $\displayranking_2$ , i.e., $\displayranking = \displayranking_2 \rightarrow c(\doc_m) = 0$, thus the expectation for $\doc_m$ only has to consider $\displayranking_1$:
\begin{equation}
\begin{split}
\mbox{}\hspace*{-1mm}
\mathbb{E}_{o, \displayranking}&\Big[
\frac{c(\doc_m) \cdot \lambda\big(\doc_m \mid \ranking \big)}{ \rho\big( o(\doc_m) = 1 \mid q, \doc_m, \displayranking, \pi \big)}
\Big] \\
& =\frac{\pi(\displayranking_1 \mid q) \cdot P\big( o(\doc_m) = 1 \mid q, r, \displayranking_1 \big)}{ \rho\big( o(\doc_m) = 1 \mid q, \doc_m, \displayranking_1, \pi \big)} \cdot \lambda\big(\doc_m \,|\, \ranking \big) \cdot r(\doc_m).
\end{split}
\end{equation}
In this situation, Condition~\ref{eq:agnosticcond} is not satisfied, and correspondingly, the policy-oblivious estimator~(Eq.~\ref{eq:obliviousestimator}) does not give an unbiased estimate:
\begin{equation}
\begin{split}
\frac{\pi(\displayranking_1 \mid q) \cdot P\big( o(\doc_m) = 1 \mid  q, r, \displayranking_1 \big)}{
P\big( o(\doc_m) = 1 \mid  q, r, \displayranking_1 \big)
} \cdot \lambda\big(\doc_m \mid  \ranking \big) \cdot r(\doc_m) \\
 < \lambda\big(\doc_m \mid  \ranking \big) \cdot r(\doc_m).
\end{split}
\end{equation}
Since the policy-oblivious estimator $\hat{\Delta}_{\textit{oblivious}}$ only corrects for the observance probability per displayed ranking, it is unable to correct for the zero probability in $R_2$ as no clicks on $\doc_m$ can occur here.
As a result, the estimate is affected by the logging policy $\pi$: the more item selection bias $\pi$ introduces (determined by $\pi(\displayranking_1 \mid q)$) the further the estimate will deviate.
Consequently, in expectation $\hat{\Delta}_{\textit{oblivious}}$ will biasedly estimate that $\doc_n$ should be ranked higher than $\doc_m$, which is incorrect since both items are actually equally relevant.

In contrast, the policy-aware estimator $\hat{\Delta}_{\textit{aware}}$~(Eq.~\ref{eq:policyaware}) avoids this issue because its propensities are based on the logging policy $\pi$.
When calculating the probability of observance conditioned on $\pi$, $P\big(o(\doc_m) = 1 \mid q, r, \pi\big)$~(Eq.~\ref{eq:expexam}), it takes into account that there is a $\pi(\displayranking_2 \mid q)$ chance that $\doc_m$ is not displayed to the user:
\begin{equation}
\begin{split}
\frac{\pi(\displayranking_1 |\,  q) \cdot P\big( o(\doc_m) = 1 |\,  q, r, \displayranking_1 \big)}{
\pi(\displayranking_1 |\,  q) \cdot P\big( o(\doc_m) = 1 |\,  q, r, \displayranking_1 \big)
} 
\cdot{} & \lambda\big(\doc_m |\,  \ranking \big) \cdot r(\doc_m) \\
={} &\lambda\big(\doc_m |\,  \ranking \big) \cdot r(\doc_m).
\end{split}
\end{equation}
Since in this situation Condition~\ref{eq:awarecond2} is true (and therefore also Condition~\ref{eq:awarecond}), we know beforehand that in expectation the policy-aware estimator is unaffected by position and item-selection bias.

This concludes our illustrative example.
It was meant to contrast the behavior of the policy-aware and policy-oblivious estimators in two different situations.
When there is no item selection bias, i.e.,  an item is displayed in the top-$k$ of all rankings the logging policy may display, both estimators provide unbiased estimates albeit using different propensity scores.
However, when there is item selection bias. i.e., an item is not always present in the top-$k$, the policy-oblivious estimator $\hat{\Delta}_{\textit{oblivious}}$ no longer provides an unbiased estimate, while the policy-aware estimator $\hat{\Delta}_{\textit{aware}}$ is still unbiased w.r.t. both position bias and item selection bias.

\section{Learning for Top-$k$ Metrics}
\label{sec:topkmetrics}

This section details how counterfactual \ac{LTR} can be used to optimize top-$k$ metrics, since these are the relevant metrics in top-$k$ rankings.

\subsection{Top-$k$ Metrics}

Since top-$k$ rankings only display the $k$ highest ranked items to the user, the performance of a ranker in this setting is only determined by those items.
Correspondingly, only top-$k$ metrics matter here, where items beyond rank $k$ have no effect:
\begin{align}
\lambda^\textit{metric@k}\big(\doc \mid \ranking \big) &=
\begin{cases}
\lambda^\textit{metric}\big(\doc \mid \ranking \big), & \text{if rank}( \doc \mid \ranking) \leq k, \\
0, & \text{if rank}\big(\doc\mid \ranking \big) > k. \\
\end{cases}
\end{align}
These metrics are commonly used in \ac{LTR} since, usually, performance gains in the top of a ranking are the most important for the user experience.
For instance, NDCG@$k$, which is the normalized version of DCG@$k$, is often used:
\begin{align}
\mbox{}\hspace*{-1.5mm}
\lambda^\textit{DCG@k}\big(\doc\,|\, \ranking \big) &\!=\!
\begin{cases}
-\log_2\big(1 \,{+}\, \text{rank}(\doc\,|\, \ranking)\big)^{-1},\hspace*{-2mm}\mbox{} & \text{if rank}(\doc\,|\,\ranking) \,{\leq}\, k, \\
0, & \text{if rank}\big(\doc\,|\, \ranking \big) \,{>}\, k.
\end{cases}
\hspace*{-2mm}\mbox{}
\end{align}
Generally in \ac{LTR}, DCG is optimized in order to maximize NDCG~\citep{wang2018lambdaloss, burges2010ranknet}.
In unbiased \ac{LTR} it is not trivial to estimate the normalization factor for NDCG, further motivating the optimization of DCG instead of NDCG~\citep{agarwal2019counterfactual, carterette2018offline}.

Importantly, top-$k$ metrics bring two main challenges for \ac{LTR}.
First, the \text{rank} function is not differentiable, a problem for almost every \ac{LTR} metric~\citep{wang2018lambdaloss, liu2009learning}.
Second,
changes in a ranking beyond position $k$ do not affect the metric's value thus resulting in zero-gradients.
The first problem has been addressed in existing \ac{LTR} methods, we will now propose adaptations of these methods that address the second issue as well.

\subsection{Monotonic Upper Bounding}
\label{section:monotonic upper bounding}
A common approach for enabling optimization of ranking methods, is by finding lower or upper bounds that can be minimized or maximized, respectively.
For instance, similar to a hinge loss, the \text{rank} function can be upper bounded by a maximum over score differences~\citep{joachims2002optimizing, joachims2017unbiased}.
Let $s$ be the scoring function used to rank (in descending order), then:
\begin{align}
\text{rank}\big(\doc\mid \ranking \big) \leq \sum_{\doc' \in \ranking} \max\Big(1 - \big(s(\doc) - s(\doc')\big), 0\Big). \label{eq:linearupperbound}
\end{align}
Alternatively, the logistic function is also a popular choice~\citep{wang2018lambdaloss}:
\begin{align}
\text{rank}\big(\doc\mid \ranking \big) \leq \sum_{\doc' \in \ranking} \log_2\Big(1 + e^{s(\doc') - s(\doc)}\Big).\label{eq:logupperbound}
\end{align}
Minimizing one of these differentiable upper bounds will directly minimize an upper bound on the ARP metric (Eq.~\ref{eq:ARP}).

Furthermore, \citeauthor{agarwal2019counterfactual}~\citep{agarwal2019counterfactual} showed that this approach can be extended to any metric based on a monotonically decreasing function.
For instance, if $\overline{\text{rank}}\big(\doc\mid \ranking \big)$ is an upper bound on the $\text{rank}\big(\doc\mid \ranking \big)$ function, then the following is an upper bound on the DCG loss (Eq.~\ref{eq:DCG}):
\begin{align}
\begin{split}
\lambda^\textit{DCG}\big(\doc\mid \ranking \big) &\leq -\log_2\big(1 + \overline{\text{rank}}(\doc\mid \ranking )\big)^{-1}  \\
&= \hat{\lambda}^\textit{DCG}\big(\doc\mid \ranking \big). 
\end{split}
\end{align}
More generally, let $\alpha$ be a monotonically decreasing function. 
A loss based on $\alpha$ is always upper bounded by:
\begin{align}
\begin{split}
\lambda^\alpha\big(\doc\mid \ranking \big) 
&= -\alpha\big(\text{rank}(\doc\mid \ranking)\big) 
\\ 
&\leq -\alpha\big(\overline{\text{rank}}(\doc\mid \ranking )\big)
= \hat{\lambda}^\alpha\big(\doc\mid \ranking \big).
\end{split}
\end{align}
Though appropriate for many standard ranking metrics, $\hat{\lambda}^\alpha$ is not an upper bound for top-$k$ metric losses.
To understand this, consider that an item beyond rank $k$ may still receive a negative score from $\hat{\lambda}^\alpha$,
for instance, for the DCG upper bound: $\text{rank}\big(\doc\mid \ranking \big) > k \rightarrow \hat{\lambda}^\textit{DCG}\big(\doc\mid \ranking \big) < 0$.
As a result, this is not an upper bound for a DCG@$k$ based loss.

We propose a modification of the $\hat{\lambda}^\alpha$ function to provide an upper bound for top-$k$ metric losses, by simply giving a positive penalty to items beyond rank $k$:
\begin{align}
\mbox{}\hspace*{-2mm}
\hat{\lambda}^{\alpha\textit{@k}}\big(\doc\mid \ranking \big) &= -\alpha\big(\overline{\text{rank}}(\doc\mid \ranking)\big) + \mathds{1}\big[\text{rank}(\doc\mid \ranking) > k\big] \cdot \alpha(k).
 \label{eq:kmonotonic}
\end{align}
The resulting function is an upper bound on top-$k$ metric losses based on a monotonic function: $\lambda^{\alpha\textit{@k}}\big(\doc\mid \ranking \big) \leq \hat{\lambda}^{\alpha\textit{@k}}\big(\doc\mid \ranking \big)$.
The main difference with $\hat{\lambda}^\alpha$ is that items beyond rank $k$ acquire a positive score from $\lambda^{\alpha\textit{@k}}$, thus providing an upper bound on the actual metric loss.
Interestingly, the gradient of $\hat{\lambda}^{\alpha\textit{@k}}$ w.r.t. the scoring function $s$ is the same as that of $\hat{\lambda}^{\alpha}$.\footnote{We consider the indicator function to never have a non-zero gradient.}
Therefore, the gradient of either function optimizes an upper bound on  $\lambda^{\alpha\textit{@k}}$ top-$k$ metric losses, while only $\hat{\lambda}^{\alpha\textit{@k}}$ provides an actual upper bound.

While this monotonic function-based approach is simple, it is unclear how coarse these upper bounds are.
In particular, some upper bounds on the rank function (e.g., Eq.~\ref{eq:linearupperbound}) can provide gross overestimations. 
As a result, these upper bounds on ranking metric losses may be very far removed from their actual values.

\subsection{Lambda-Based Losses for Counterfactual top-$k$ \ac{LTR}}
\label{section:lambda-based losses}
Many supervised \ac{LTR} approaches, such as  the well-known LambdaRank and subsequent LambdaMART methods~\citep{burges2010ranknet}, are based on \ac{EM} procedures~\citep{dempster1977maximum}.
Recently, \citeauthor{wang2018lambdaloss}~\citep{wang2018lambdaloss} introduced the Lambda\-Loss framework, which provides a theoretical way to prove that a method optimizes a lower bound on a ranking metric.
Subsequently, it was used to prove that Lambda\-MART optimizes such a bound on \ac{DCG}, similarly it was also used to introduce the novel Lambda\-Loss method which provides an even tighter bound on \ac{DCG}.
In this section, we will show that the Lambda\-Loss framework can be used to find proven bounds on counterfactual \ac{LTR} losses and top-$k$ metrics.
Since Lambda\-Loss is considered state-of-the-art in supervised \ac{LTR}, making its framework applicable to counterfactual \ac{LTR} could potentially provide competitive performance.
Additionally, adapting the Lambda\-Loss framework to top-$k$ metrics further expands its applicability.

The Lambda\-Loss framework and its \ac{EM}-optimization approach work for metrics that can be expressed in item-based gains, $G(\doc_n \mid q,r)$, and discounts based on position, $D\big(\text{rank}(\doc_n \mid \ranking)\big)$; for brevity we use the shorter $G_n$ and $D_n$, respectively, resulting in:
\begin{align}
\mbox{}
\hspace*{-2mm}
\Delta\big(\ranking \,|\, q, r \big)
=
\sum_{\doc_n \in \ranking} G(\doc_n | \, q, r) \cdot D\big(\text{rank}(\doc_n | \, \ranking)\big)
=
\sum_{n=1}^{|\ranking|} G_n \cdot D_n.
\end{align}
For simplicity of notation, we choose indexes so that: $n = \text{rank}(\doc_n \mid \ranking)$, thus $D_n$ is always the discount for the rank $n$.
Then, we differ from the existing Lambda\-Loss framework by allowing the discounts to be zero ($\forall n \, D_n \geq 0$), thus also accounting for top-$k$ metrics.
Furthermore, items at the first rank are not discounted or the metric can be scaled so that $D_1 = 1$.
Additionally, higher ranked items should be discounted less or equally: $n > m \rightarrow D_n \leq D_m$.
Most ranking metrics meet these criteria; for instance, $G_n$ and $D_n$ can be chosen to match \ac{ARP} or \ac{DCG}.
Importantly, our adaption also allows $\Delta$ to match top-$k$ metrics such as \ac{DCG}$@k$ or Precision$@k$.

In order to apply the Lambda\-Loss framework to counterfactual \ac{LTR}, we consider a general inverse-propensity-scored estimator:
\begin{align}
\hat{\Delta}_{\textit{IPS}}(\ranking \mid q, c, \cdot)
&= \sum_{\doc_n:c(\doc_n) = 1} \frac{\lambda(\doc_n\mid \ranking)}{\rho\big( o(\doc_n) = 1 \mid q, r, \displayranking, \pi \big) },
\label{eq:generalestimator}
\end{align}
where the propensity function $\rho$ can match either the policy-obli\-vious (Eq.~\ref{eq:obliviousestimator}) or the policy-aware (Eq.~\ref{eq:policyaware}) estimator.
By choosing 
\begin{align}
G_n = \frac{1}{\rho\big( o(\doc_n) = 1 \mid q, r, \displayranking, \pi  \big) }\text{ and }D_n = \lambda(\doc_n\mid \ranking),
\label{eq:gaindiscountchoice}
\end{align}
the estimator can be described in terms of gains and discounts.
In contrast, in the existing Lambda\-Loss framework~\citep{wang2018lambdaloss} gains are based on item relevance. 
For counterfactual top-$k$ \ac{LTR}, we have designed Eq.~\ref{eq:gaindiscountchoice} so that gains are based on the propensity scores of observed clicks, and the discounts can have zero values.

The \ac{EM}-optimization procedure alternates between an expectation step and a maximization step.
In our case, the expectation step sets the discount values $D_n$ according to the current ranking $\ranking$ of the scoring function $s$.
Then the maximization step updates $s$ to optimize the ranking model.
Following the Lambda\-Loss framework~\citep{wang2018lambdaloss}, we derive a slightly different loss.
With the delta function:
\begin{align}
\delta_{nm} = D_{|n - m|} - D_{|n - m| + 1},
\end{align}
our differentiable \emph{counterfactual loss} becomes:
\begin{align}
\sum_{G_n > G_m} -\log_2 \left( \left (\frac{1}{1 + e^{s(\doc_m) - s(\doc_n)}} \right )^{\delta_{nm} \cdot |G_n - G_m|} \right). \label{eq:embound}
\end{align}
The changes we made do not change the validity of the proof provided in the original Lambda\-Loss paper~\citep{wang2018lambdaloss}.
Therefore, the counterfactual loss (Eq.~\ref{eq:embound}) can be proven to optimize a lower bound on counterfactual estimates of top-$k$ metrics.

Finally, in the same way the LambdaLoss framework can also be used to derive counterfactual variants of other supervised \ac{LTR} losses/methods such as LambdaRank or LamdbaMART.
Unlike previous work that also attempted to find a counterfactual lambda-based method by introducing a pairwise-based estimator~\cite{hu2019unbiased}, our approach is compatible with the prevalent counterfactual approach since it uses the same estimator based on single-document propensities~\citep{wang2016learning, joachims2017unbiased, agarwal2019addressing, ai2018unbiased, wang2018position, agarwal2019estimating}.
Our approach suggests that the division between supervised and counterfactual \ac{LTR} methods may disappear in the future, as a state-of-the-art supervised \ac{LTR} method can now be applied to the state-of-the-art counterfactual \ac{LTR} estimators.

\subsection{Unbiased Loss Selection}
So far we have introduced two counterfactual \ac{LTR} approaches that are proven to optimize lower bounds on top-$k$ metrics: with monotonic functions~(Section~\ref{section:monotonic upper bounding}) and through the LambdaLoss framework~(Section~\ref{section:lambda-based losses}).
To the best of our knowledge, we are the first to introduce theoretically proven lower bounds for top-$k$ \ac{LTR} metrics.
Nevertheless, previous work has also attempted to optimize top-$k$ metrics, albeit through heuristic methods.
Notably, \citet{wang2018lambdaloss} used a truncated version of the LambdaLoss loss to optimize \ac{DCG}$@k$.
Their loss uses the discounts $D_n$ based on full-ranking \ac{DCG} but ignores item pairs outside of the top-$k$: 
\begin{align}
\begin{split}
\sum_{G_n > G_m} -\mathds{1}\big[& n \leq k \lor m \leq k\big] \cdot {}\\
& \log_2  \left( \left ( \frac{1}{1 + e^{s(\doc_m) - s(\doc_n)}} \right )^{\delta_{nm} \cdot |G_n - G_m|}  
 \right ).
\end{split}  \label{eq:truncembound}
\end{align}
While empirical results motivate its usage, there is no known theoretical justification for this loss, and thus it is considered a heuristic.

This leaves us with a choice between two theoretically motivated counterfactual \ac{LTR} approaches for optimizing top-$k$ metrics (Eq.~\ref{eq:kmonotonic}~and~\ref{eq:embound}) and an empirically motivated heuristic (Eq.~\ref{eq:truncembound}).
We propose a pragmatic solution by recognizing that counterfactual estimators can unbiasedly evaluate top-$k$ metrics.
Therefore, in practice one can optimize several ranking models using various approaches, and subsequently, estimate which resulting model provides the best performance.
Thus, using counterfactual evaluation to select from resulting models is an unbiased method to choose between the available counterfactual \ac{LTR} approaches.

\begin{figure*}[t]
\centering
\vspace{-0.7\baselineskip}
\begin{tabular}{r r c }
\multicolumn{1}{c}{ \footnotesize {Yahoo! Webscope}}
&
\multicolumn{1}{c}{ \footnotesize {MSLR-WEB30k}}
\\
\includegraphics[scale=0.32]{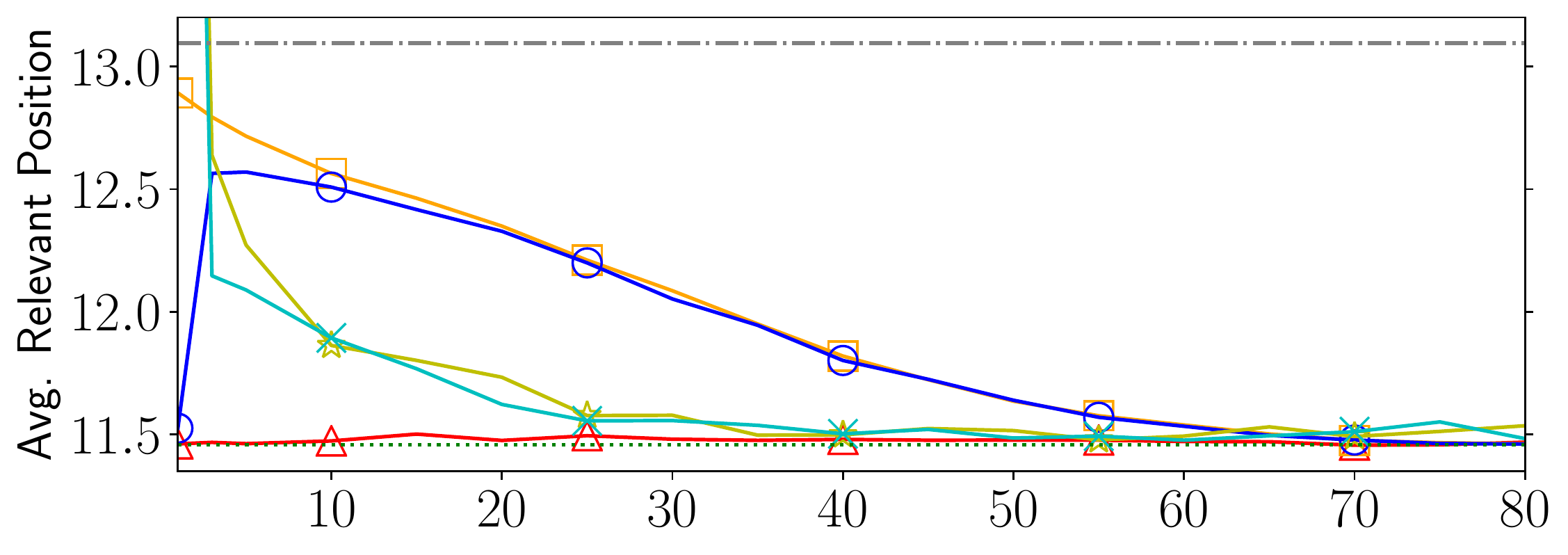} &
\includegraphics[scale=0.32]{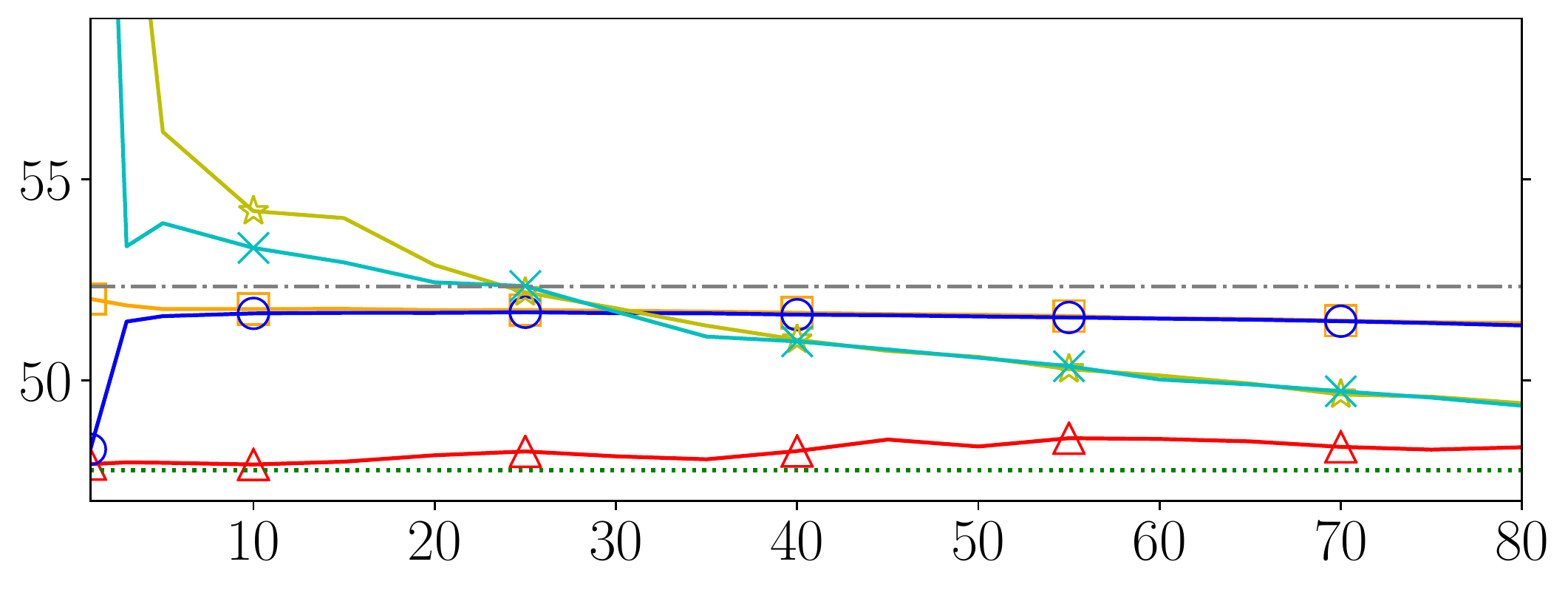} &
\multirow{2}{*}[6.3em]{ \includegraphics[scale=.29]{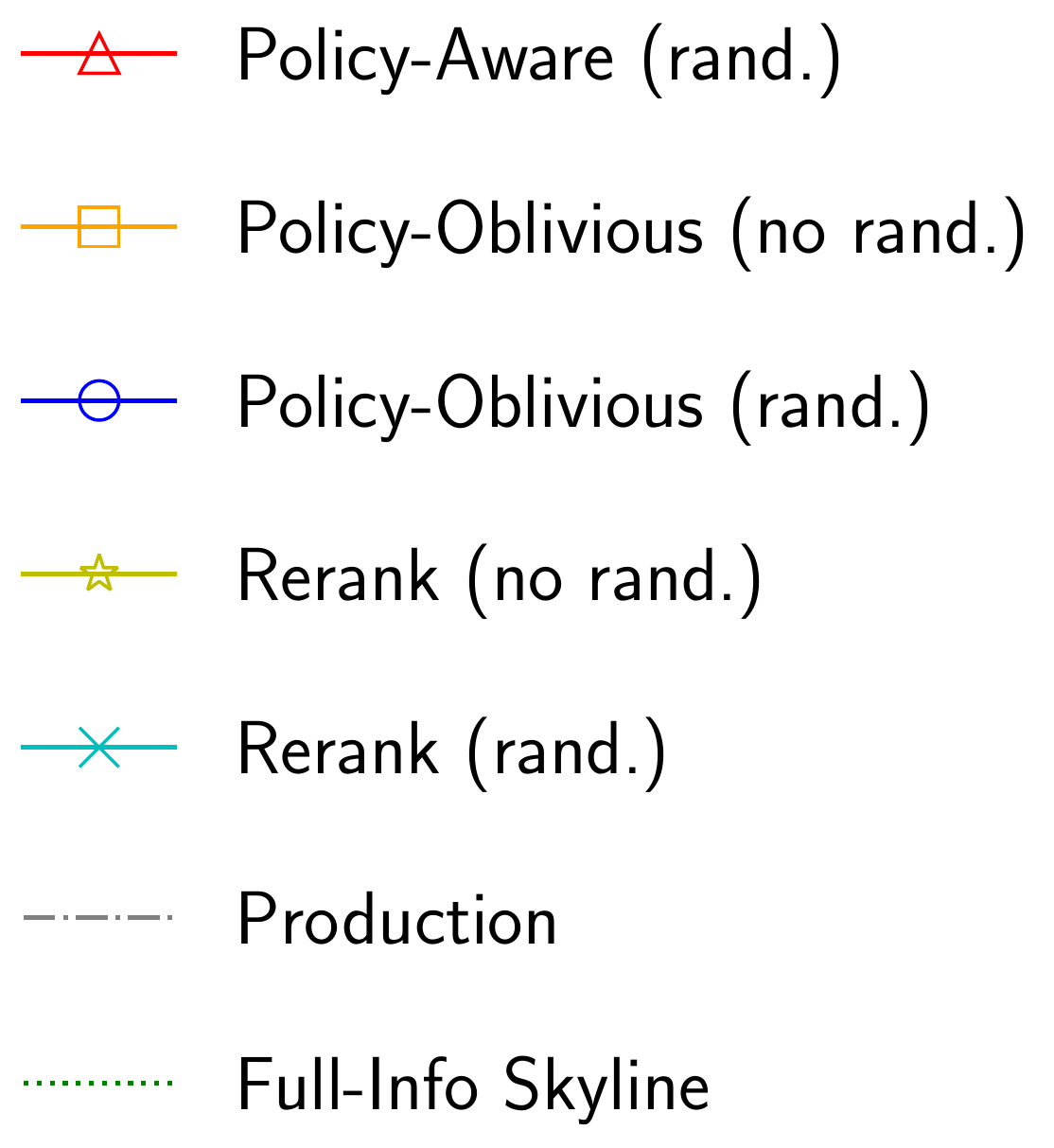}}
\vspace{-0.3em}
\\
\includegraphics[scale=0.32]{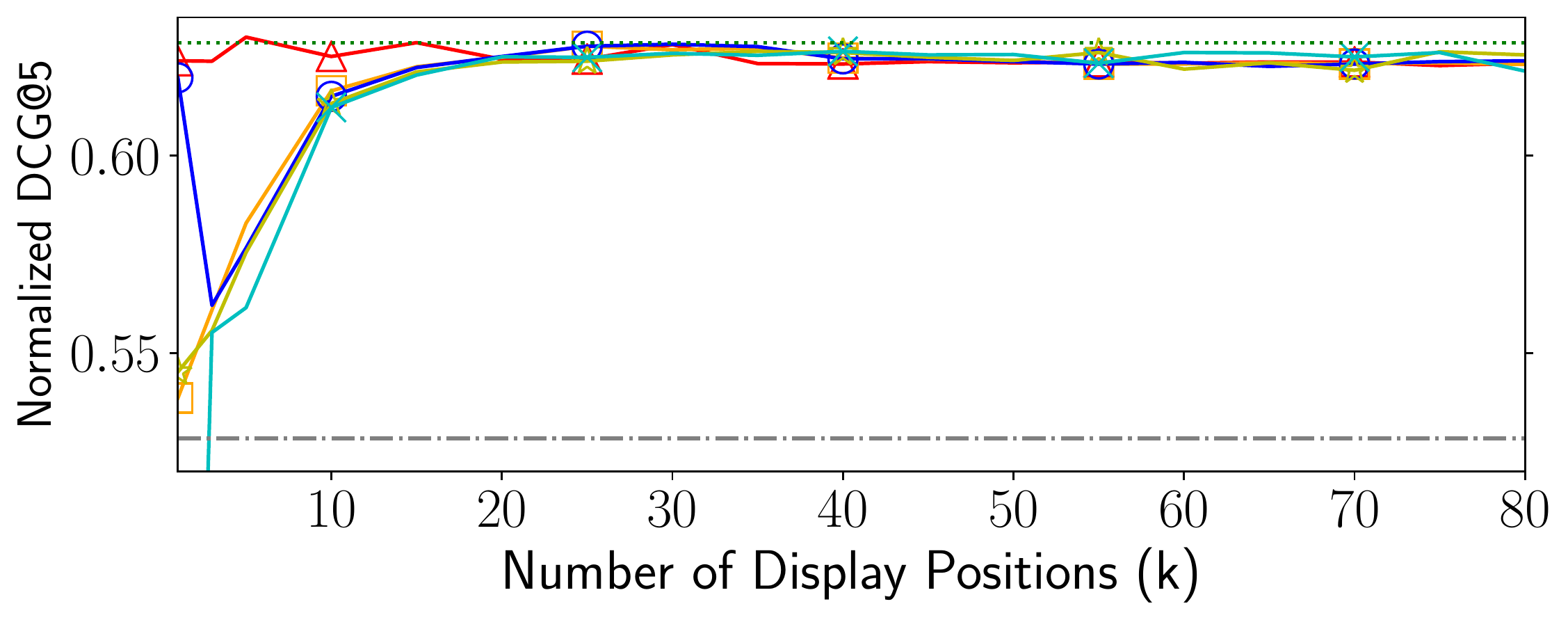} &
\includegraphics[scale=0.32]{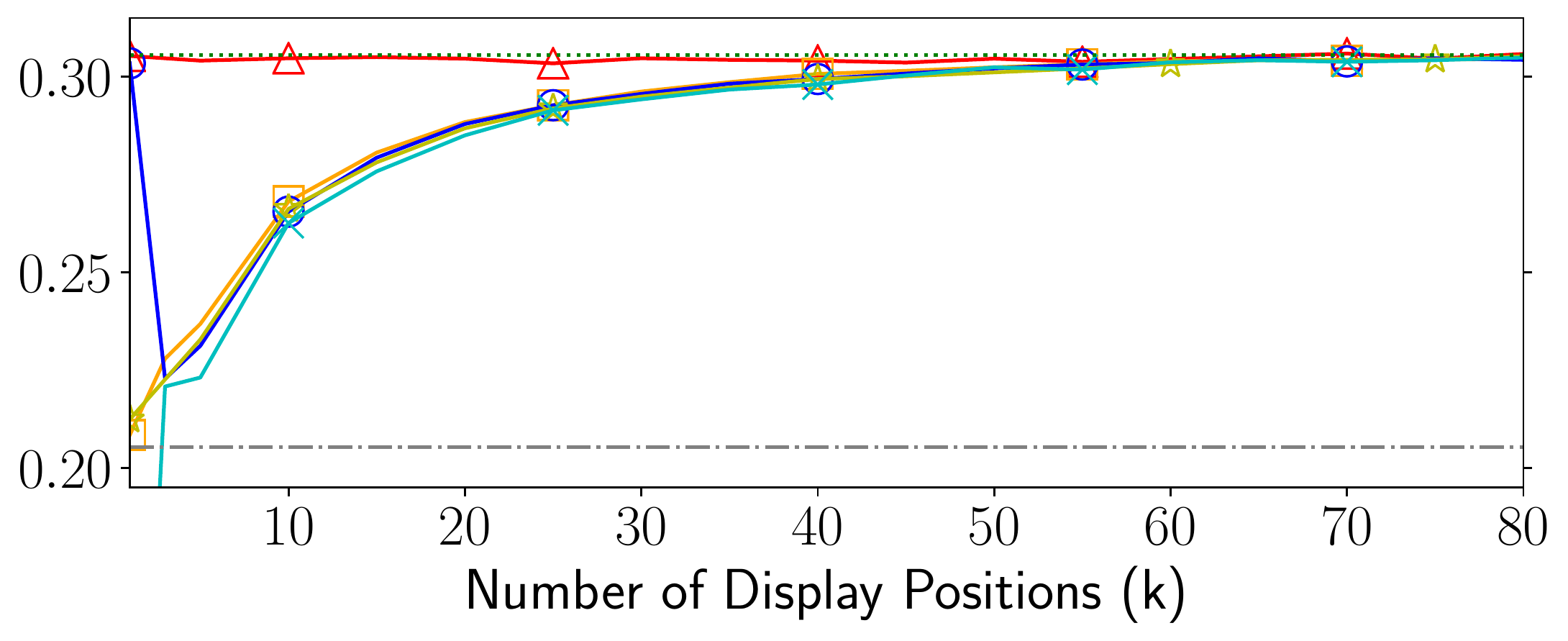} 
\end{tabular}
\caption{
The effect of item selection bias on different estimators.
Optimization on $10^8$ clicks simulated on top-$k$ rankings with varying number of display positions ($k$), with and without randomization (for each datapoint $10^8$ clicks were simulated independently).
The top row optimizes the average relevance position through the linear upper bound (Eq.~\ref{eq:linearupperbound});
the bottom row optimizes DCG$@5$ using the truncated LambdaLoss (Eq.~\ref{eq:truncembound}).
Left: results on the \emph{Yahoo} dataset; right: on the \emph{MSLR} dataset. 
}
\label{fig:effectk}
\end{figure*}

\section{Experimental Setup}
We follow the standard setup in unbiased \ac{LTR}~\citep{joachims2017unbiased, ai2018unbiased, carterette2018offline, jagerman2019comparison} and perform semi-synthetic experiments: queries and items are based on datasets of commercial search engines and interactions are simulated using probabilistic click models.

\subsection{Datasets}
We use the queries and documents from two of the largest publicly available \ac{LTR} datasets: MLSR-WEB30K~\citep{qin2013introducing} and Yahoo! Webscope~\citep{Chapelle2011}.
 Each was created by a commercial search engine and contains a set of queries with corresponding preselected document sets.
 Query-document pairs are represented by feature vectors and five-grade relevance annotations ranging from not relevant (0) to perfectly relevant (4).
 In order to binarize the relevancy, we only consider the two highest relevance grades as relevant.
 The MSLR dataset contains \numprint{30000} queries with on average 125 preselected documents per query, and encodes query-document pairs in 136 features.
 The Yahoo dataset has \numprint{29921} queries and on average 24 documents per query encoded in 700 features. 
Presumably, learning from top-$k$ feedback is harder as $k$ becomes a smaller percentage of the number of items.
Thus, we expect the MSLR dataset with more documents per query to pose a more difficult problem.
 
\subsection{Simulating Top-$k$ Settings}

The setting we simulate is one where interactions are gathered using a non-optimal but decent production ranker.
We follow existing work~\citep{joachims2017unbiased, ai2018unbiased, jagerman2019comparison} and use supervised optimization for the \ac{ARP} metric on 1\% of the training data.
The resulting model simulates a real-world production ranker since it is much better than a random initialization but leaves enough room for improvement~\citep{joachims2017unbiased}.

We then simulate user-issued queries by uniformly sampling from the training partition of the dataset.
Subsequently, for each query the production ranker ranks the documents preselected by the dataset.
Depending on the experimental run that we consider, randomization is performed on the resulting rankings.
In order for the policy-aware estimator to be unbiased, every relevant document needs a chance of appearing in the top-$k$ (Condition~\ref{eq:awarecond2}).
Since in a realistic setting relevancy is unknown, we choose to give every document a non-zero probability of appearing in the top-$k$.
Our randomization policy takes the ranking of the production ranker and leaves the first $k-1$ documents unchanged
but the document at position $k$ is selected by sampling uniformly from the remaining documents.
The result is a minimally invasive randomized top-$k$ ranking since most of the ranking is unchanged and the placement of the sampled documents is limited to the least important position.

We note that many other logging policies could be applied (see Condition~\ref{eq:awarecond2}), e.g., an alternative policy could insert sampled documents at random ranks for less obvious randomization.
Unfortunately, a full exploration of the effect of using different logging policies is beyond the scope of this work.

Clicks are simulated on the resulting ranking $\displayranking$ according to position bias and document relevance.
Top-$k$ position bias is modelled through the probability of observance, as follows:
\begin{align}
P\big(o(\doc) = 1 \mid q, r, \displayranking \big) &=
\begin{cases}
\text{rank}(\doc\mid \displayranking)^{-1},
& \text{if rank}(\doc\mid \displayranking) \leq k,
\\
0,
& \text{if rank}(\doc\mid \displayranking) > k.
\end{cases}
\end{align}
The randomization policy results in the following examination probabilities w.r.t. the logging policy (cf.\ Eq.~\ref{eq:expexam}):
\begin{align}
\begin{split}
P\big(o(\doc) = 1 \mid &q, r, \pi \big) \\
&=
\begin{cases}
\text{rank}(\doc\mid \displayranking)^{-1}, & \text{if rank}(\doc\mid \displayranking) < k, \\
\big(k \cdot (|\displayranking| - k + 1)\big)^{-1},  & \text{if rank}(\doc\mid\displayranking) \geq k. 
\end{cases}
\end{split}
\end{align}
The probability of a click is conditioned on the relevance of the document according to the dataset:
\begin{align}
P\big(c(\doc) = 1 \mid q, r, \displayranking, o\big) &=
\begin{cases}
1,
& \text{if } r(\doc) = 1 \land o(\doc) = 1,
\\
0.1,
& \text{if } r(\doc) = 0 \land o(\doc) = 1,
\\
0,
& \text{if } o(\doc) = 0.
\end{cases}
\end{align}
Note that our previous assumption that clicks only take place on relevant items (Section~\ref{section:counterfactualLTR}) is not true in our experiments.

Optimization is performed on training clicks simulated on the training partition of the dataset.
Hyperparameter tuning is done by estimating performance on (unclipped) validation clicks simulated on the validation partition; the number of validation clicks is always 15\% of the number of training clicks.
Lastly, evaluation metrics are calculated on the test partition using the dataset labels.

\begin{figure*}[t]
\vspace{-0.7\baselineskip}
\centering
\begin{tabular}{r r l }
\multicolumn{1}{c}{ \footnotesize  {Yahoo! Webscope}}
&
\multicolumn{1}{c}{ \footnotesize  {MSLR-WEB30k}}
\\
\includegraphics[scale=0.32]{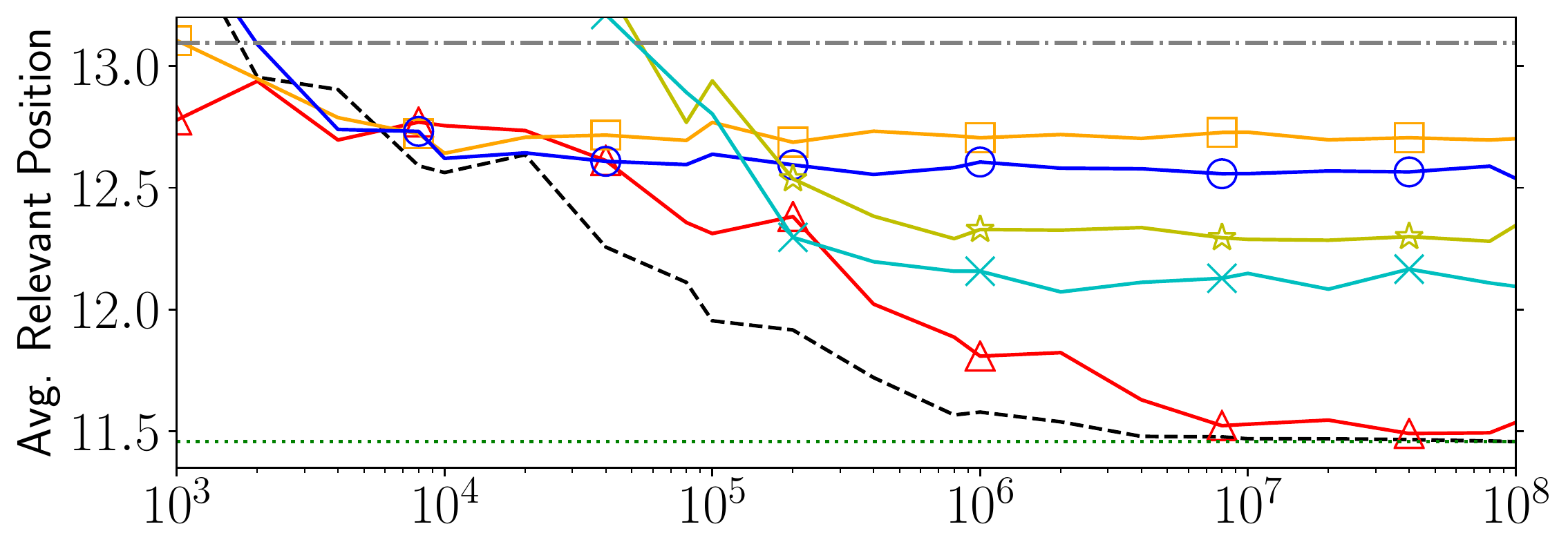} &
\includegraphics[scale=0.32]{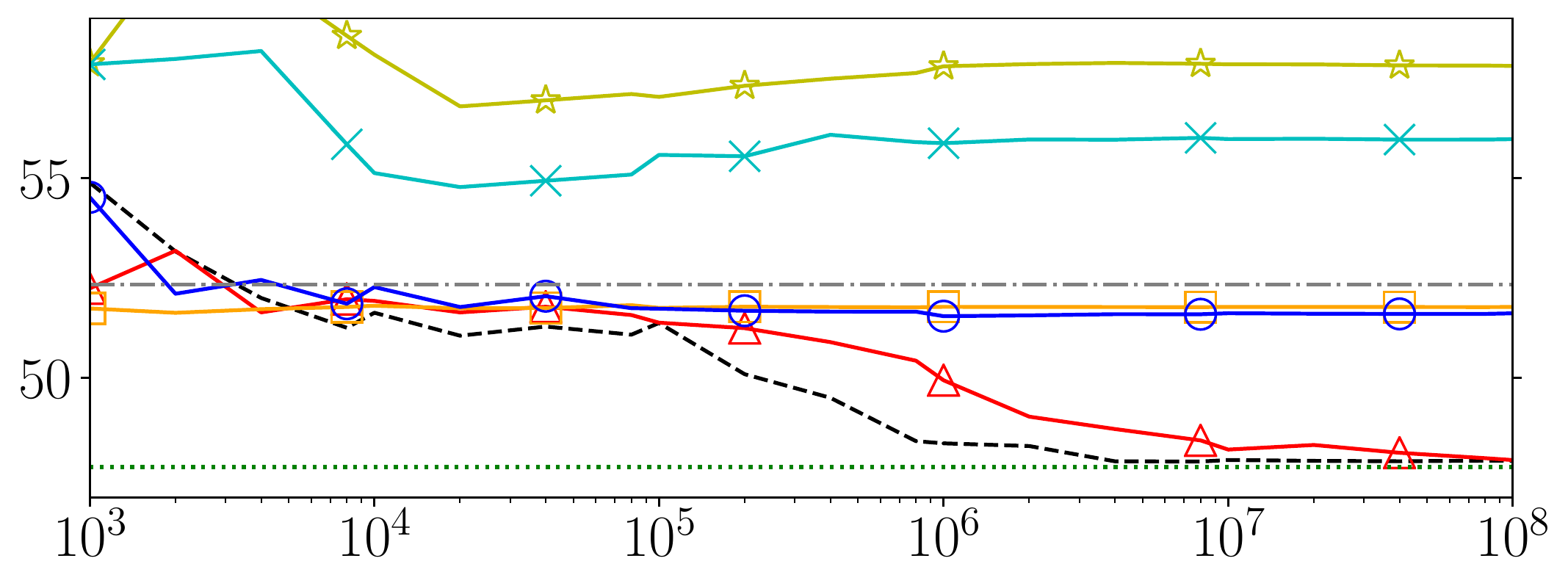} &
\multirow{2}{*}[6.5em]{ \includegraphics[scale=.27]{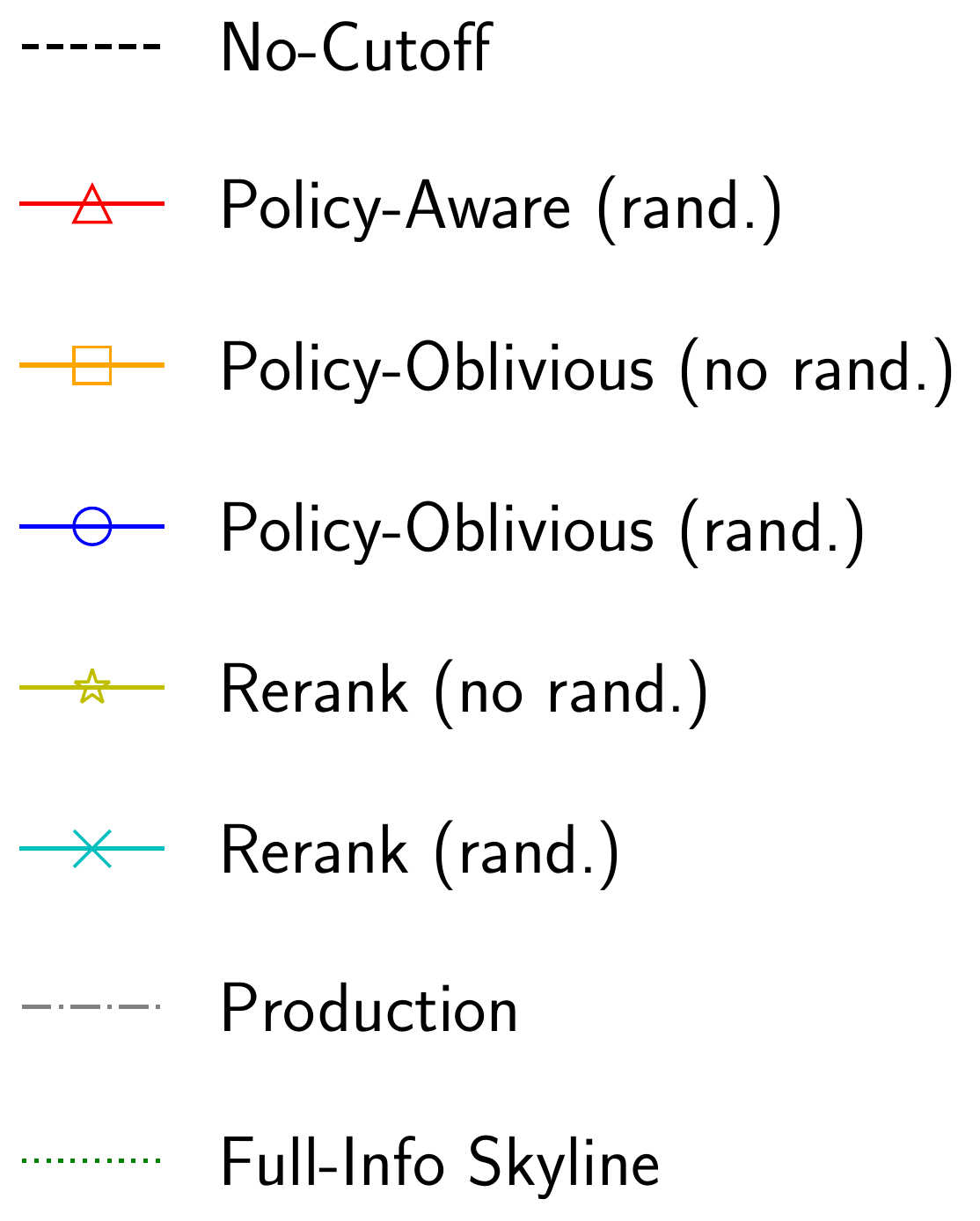}} 
\vspace{-0.3em}
\\
\includegraphics[scale=0.32]{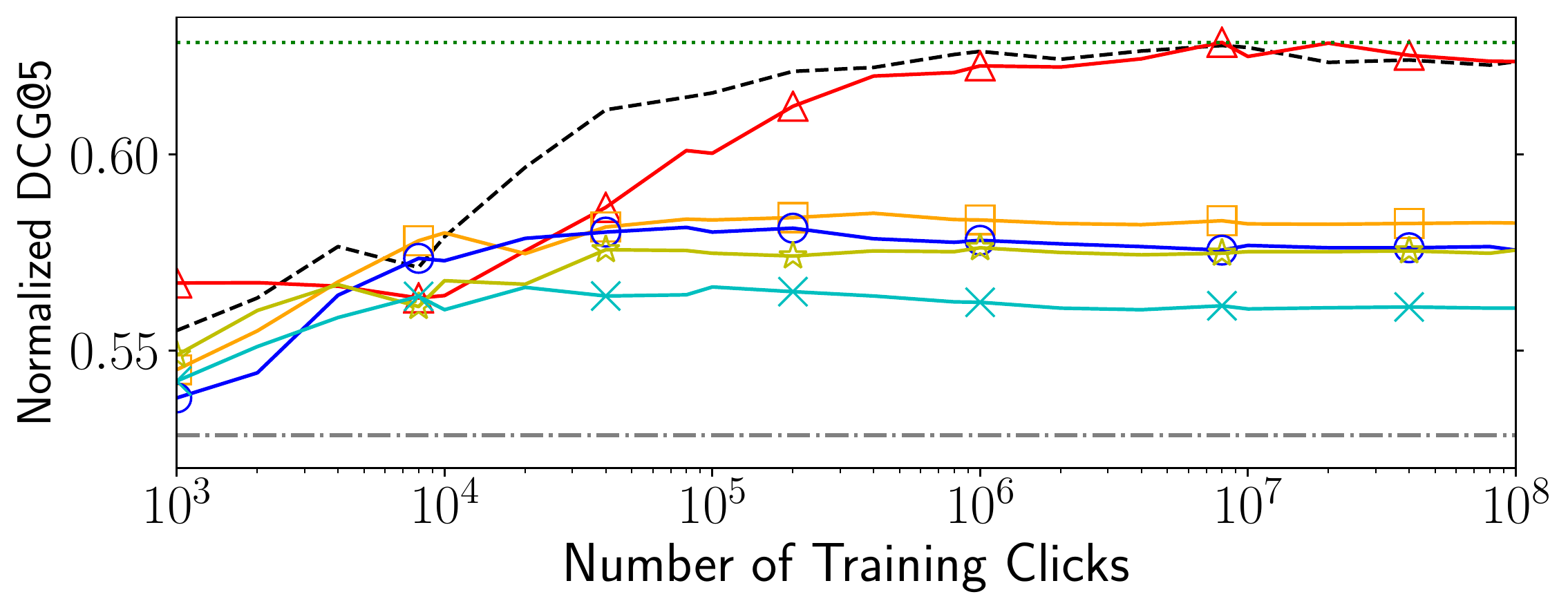} &
\includegraphics[scale=0.32]{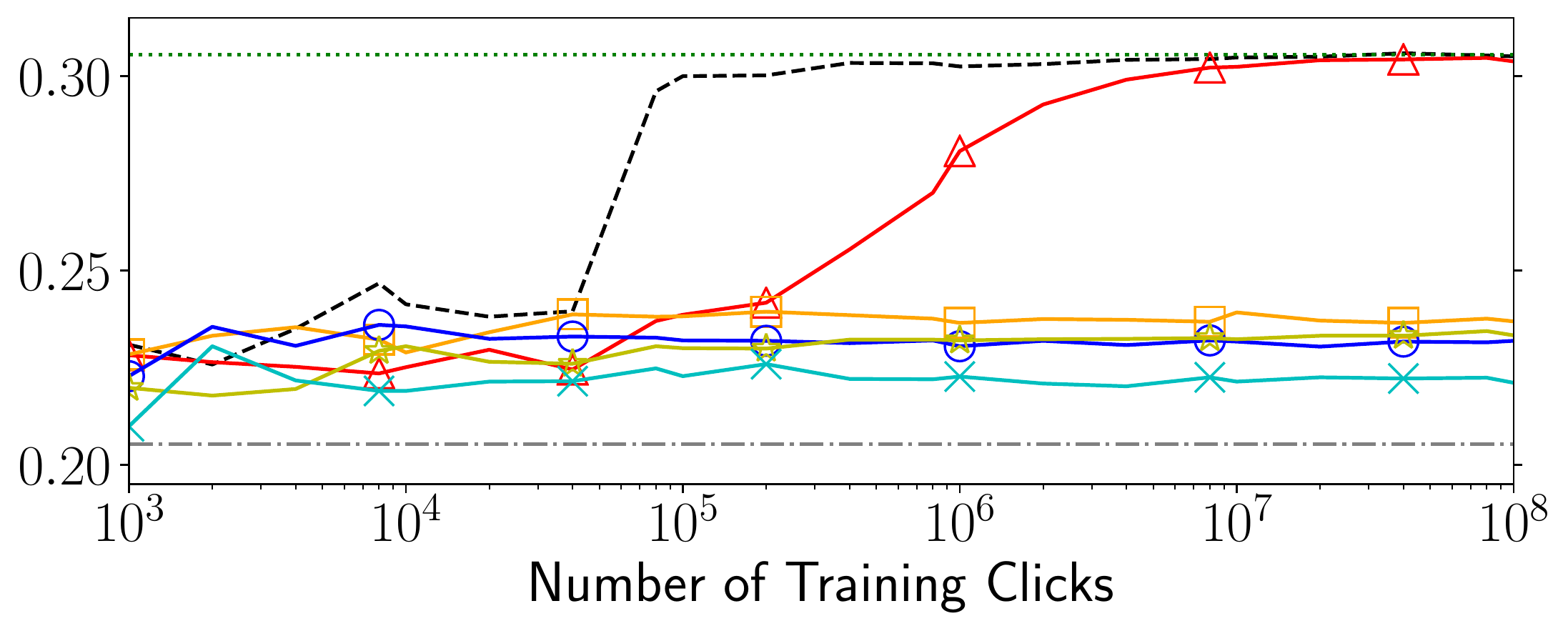} 
\end{tabular}
\caption{
Performance of different estimators learning from different numbers of clicks simulated on top-5 rankings, with and without randomization.
The top row optimizes the average relevance position through the linear upper bound (Eq.~\ref{eq:linearupperbound});
the bottom row optimizes DCG$@5$ using the truncated LambdaLoss (Eq.~\ref{eq:truncembound}).
}
\label{fig:cutoffbaseline}
\end{figure*}

\subsection{Experimental Runs}

In order to evaluate the performance of the policy-aware estimator (Eq.~\ref{eq:policyaware}) and the effect of item selection bias, we compare with the following baselines: 
\begin{enumerate*}[label=(\roman*)]
\item The policy-oblivious estimator (Eq.~\ref{eq:obliviousestimator}).
In our setting, where the examination probabilities are known beforehand, the policy-oblivious estimator also represents methods that jointly estimate these probabilities while performing \ac{LTR}, i.e., the following methods reduce to this estimator if the examination probabilities are given:~\citep{wang2016learning, joachims2017unbiased, agarwal2019addressing, ai2018unbiased}.
\item A rerank estimator, an adaption of the policy-oblivious estimator.
During optimization the rerank estimator applies the policy-oblivious estimator but limits the document set of an interaction $i$ to the $k$ displayed items $\ranking_i = \{\doc \mid \text{rank}(\doc|\displayranking_i) \leq k\}$ (cf.\ Eq.~\ref{eq:obliviousestimator}).
Thus, it is optimized to rerank the top-$k$ of the production ranker only, but during inference it is applied to the entire document set.
\item Additionally, we evaluate performance without any cutoff $k$ or randomization; in these circumstances all three estimators (Policy-Aware, Policy-Oblivious, Rerank) are equivalent.
\item Lastly, we use supervised \ac{LTR} on the dataset labels to get a \emph{full-information skyline}, which shows the hypothetical optimal performance.
\end{enumerate*}

To evaluate the effectiveness of our proposed loss functions for optimizing top-$k$ metrics, we apply the monotonic lower bound (Eq.~\ref{eq:kmonotonic}) with a linear (Eq.~\ref{eq:linearupperbound}) and a logistic upper bound (Eq.~\ref{eq:logupperbound}).
Additionally, we apply several versions of the LamdbaLoss loss function (Eq.~\ref{eq:embound}): one that optimizes full \ac{DCG}, another that optimizes \ac{DCG}$@5$, and the heuristic truncated loss also optimizing \ac{DCG}$@5$ (Eq.~\ref{eq:truncembound}).
Lastly, we apply unbiased loss selection where we select the best-performing model based on the estimated performance on the (unclipped) validation clicks.

Optimization is done with stochastic gradient descent; to maximize computational efficiency we rewrite the loss (Eq.~\ref{eq:highlevelloss}) for a propensity scoring function $\rho$ in the following manner:
\begin{align}
\hat{\mathcal{L}} &=
\frac{1}{N} \sum^N_{i=1} \hat{\Delta}\big(\ranking_i | q_i, \displayranking_i, \pi, c_i\big)
=
\frac{1}{N} \sum^N_{i=1} \sum_{\doc : c_i(\doc) = 1} \frac{\lambda\big(\doc\mid \ranking_i \big)}{\rho\big( o_i(\doc) = 1 | q_i, r, \cdot \big)}
\nonumber
\\
&= \frac{1}{N}  \sum_{q \in \mathcal{Q}} \sum_{\doc \in \ranking_q} \left ( \sum^N_{i = 1} \frac{ \mathds{1}[q_i = q] \cdot c_i(\doc)}{\rho\big( o_i(\doc) = 1 \mid q, r, \cdot\big)} \right ) \cdot  \lambda\big(\doc \mid \ranking_q\big) 
\\
&= \frac{1}{N}  \sum_{q \in \mathcal{Q}} \sum_{\doc \in \ranking_q}  \omega_{\doc} \cdot \lambda\big(\doc \mid \ranking_q \big). \nonumber
\end{align}
After precomputing the document weights $\omega_{\doc}$,
the complexity of computing the loss is only determined by the dataset size.
This allows us to optimize over very large numbers of clicks with very limited increases in computational costs.

We optimize linear models, but our approach can be applied to any differentiable model~\citep{agarwal2019counterfactual}.
Propensity clipping~\citep{joachims2017unbiased} is applied to training clicks and never applied to the validation clicks; we also use self-normalization~\citep{swaminathan2015self}.

\section{Results and Discussion}
In this section we discuss the results of our experiments and evaluate our policy-aware estimator and the methods for top-$k$ \ac{LTR} metric optimization empirically.

\subsection{Learning under Item Selection Bias}

First we consider the question: \emph{Is the policy-aware estimator effective for unbiased counterfactual \ac{LTR} from top-$k$ feedback?}
Figure~\ref{fig:effectk} displays the performance of different approaches after optimization on $10^8$ clicks under varying values for $k$.
Both the policy-oblivious and rerank estimators are greatly affected by the item selection bias introduced by the cutoff at $k$.
On the MSLR dataset neither approach is able to get close to optimal \ac{ARP} performance, optimal \ac{DCG}$@5$ is only reached when $k >50$.
On the Yahoo dataset, the policy-oblivous approach can only approximate optimal \ac{ARP} when $k > 60$; for \ac{DCG}$@5$ it requires $k > 25$.
The rerank approach reaches optimal \ac{ARP} when $k>50$ and optimal \ac{DCG}$@5$ when $k > 20$.
Considering that on average a query in the Yahoo dataset only has 24 preselected documents, it appears that even a little item selection bias has a substantial effect on both estimators.
Furthermore, randomization appears to have a very limited positive effect on the policy-oblivious and rerank approaches.
The one exception is the policy-oblivious approach when $k=1$ where it reaches optimal performance under randomization.
Here, the randomization policy gives every item an equal probability of being presented, thus trivially removing item selection bias; additionally, there is no position bias as there is only a single position.
However, besides this trivial exception, the baseline estimators are strongly affected by item selection bias and simply logging with randomization is unable to remove the effect of item selection bias.

In contrast, the policy-aware approach is hardly affected by the choice of $k$.
It consistently approximates optimal performance in terms of \ac{ARP} and \ac{DCG}$@5$ on both datasets.
On the MSLR dataset, the policy-aware approach provides near optimal \ac{ARP} performance; however, for $k > 15$ there is a small but noticeable gap.
We suspect that this is a result of variance from click-noise and can be closed by gathering more clicks.
Across all settings, the policy-aware approach appears unaffected by the choice of $k$ and thus the effect of item selection bias.
Moreover, it consistently provides performance at least as good as the baselines; and on the Yahoo dataset it outperforms them for $k < 20$ and on the MSLR dataset outperforms them for all tested values of $k$.
We note that the randomization policy is the same for all methods; in other words, under randomization the clicks for the policy-oblivious, policy-aware and rerank approaches are acquired in the exact same way.
Thus, our results show that in order to benefit from randomization, a counterfactual \ac{LTR} method has to take its effect into account, hence only the policy-aware approach has improved performance.

Figure~\ref{fig:cutoffbaseline} displays the performance when learning from top-5 feedback while varying the number of clicks.
Here we see that the policy-oblivious approach performance is stable after $10^5$ clicks have been gathered.
The rerank approach has stable performance after $10^6$ clicks when optimized for \ac{ARP} and $10^5$ for \ac{DCG}$@5$.
Both baseline approaches show biased behavior where adding additional data does not lead to improved performance.
This confirms that their estimators are unable to deal with item selection bias.
In contrast, the policy-aware approach reaches optimal performance in all settings.
However, it appears that the policy-aware approach requires more clicks than the no-cutoff baseline; we suspect that this difference is due to variance added by the randomization and smaller propensity scores.

In conclusion, we answer our first question positively: our results show that the policy-aware approach is unbiased w.r.t.\ item selection bias and position bias.
Where all baseline approaches are affected by item selection bias even in small amounts, the policy-aware approach approximates optimal performance regardless of the cutoff value $k$.

\begin{figure*}[t]
\vspace{-0.7\baselineskip}
\centering
\begin{tabular}{r r c }
\multicolumn{1}{c}{ \footnotesize  {Yahoo! Webscope}}
&
\multicolumn{1}{c}{ \footnotesize  {MSLR-WEB30k}}
\\
\includegraphics[scale=0.31]{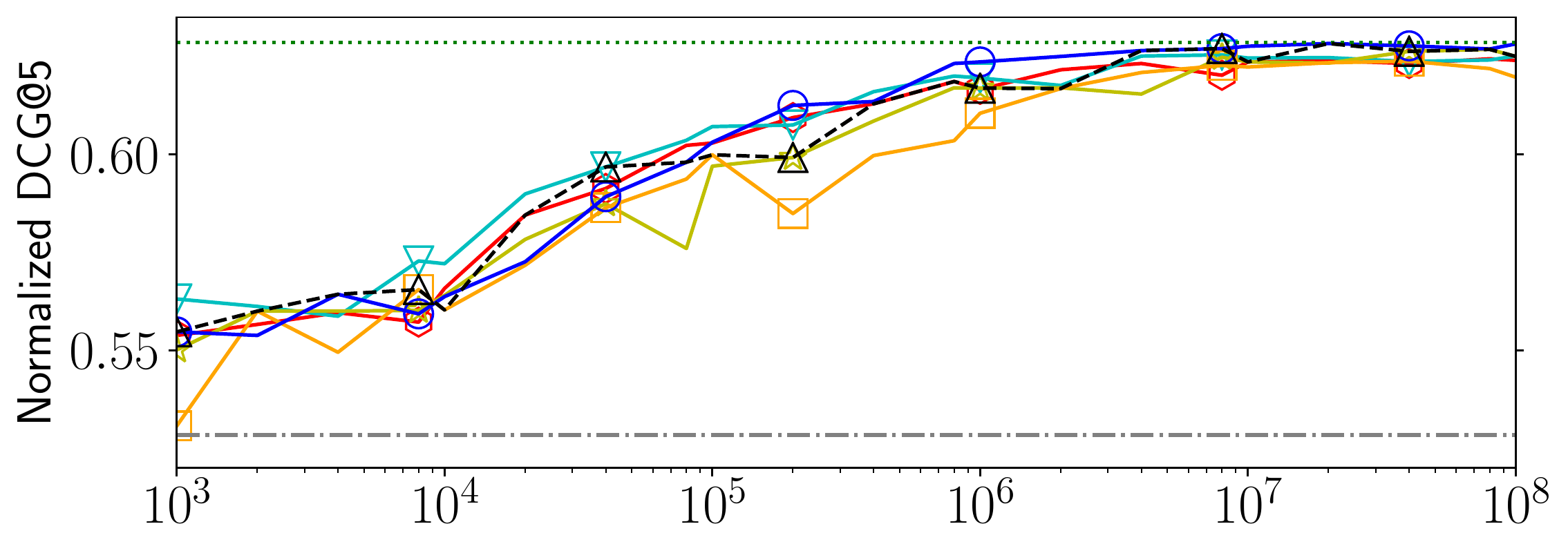} &
\includegraphics[scale=0.31]{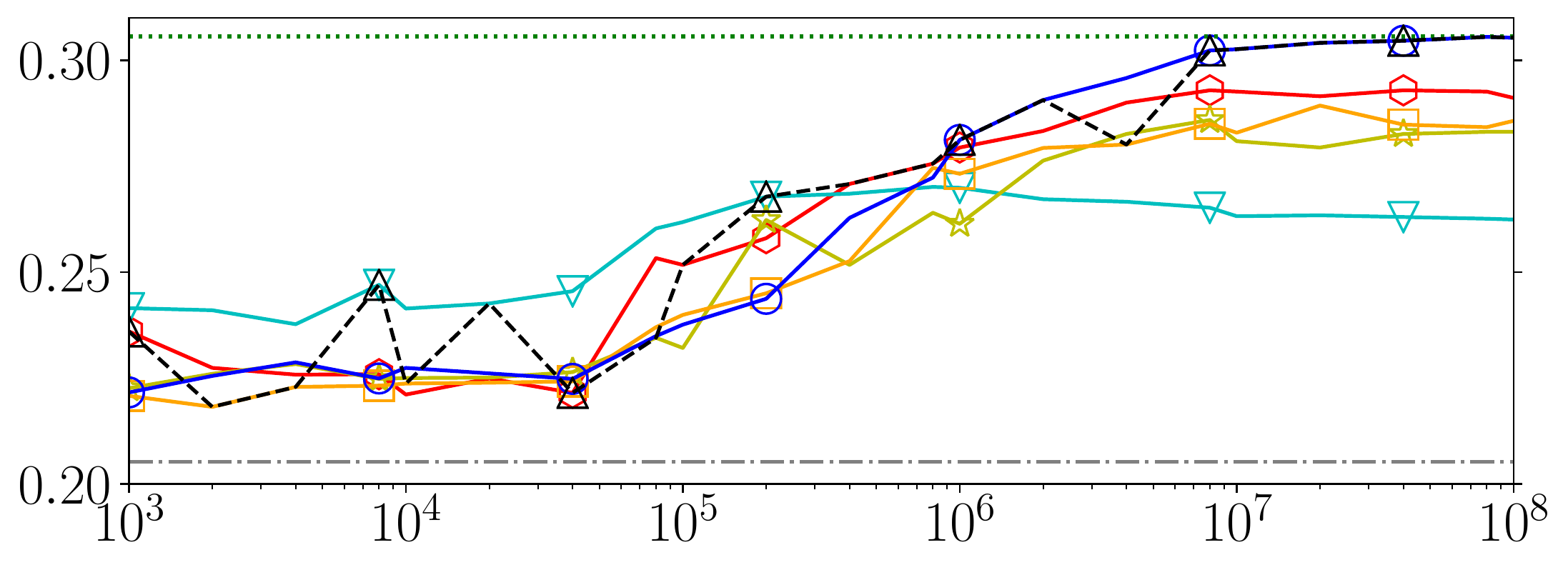} &
\multirow{2}{*}[6.5em]{ \includegraphics[scale=.27, trim=10pt 0pt 0pt 0pt]{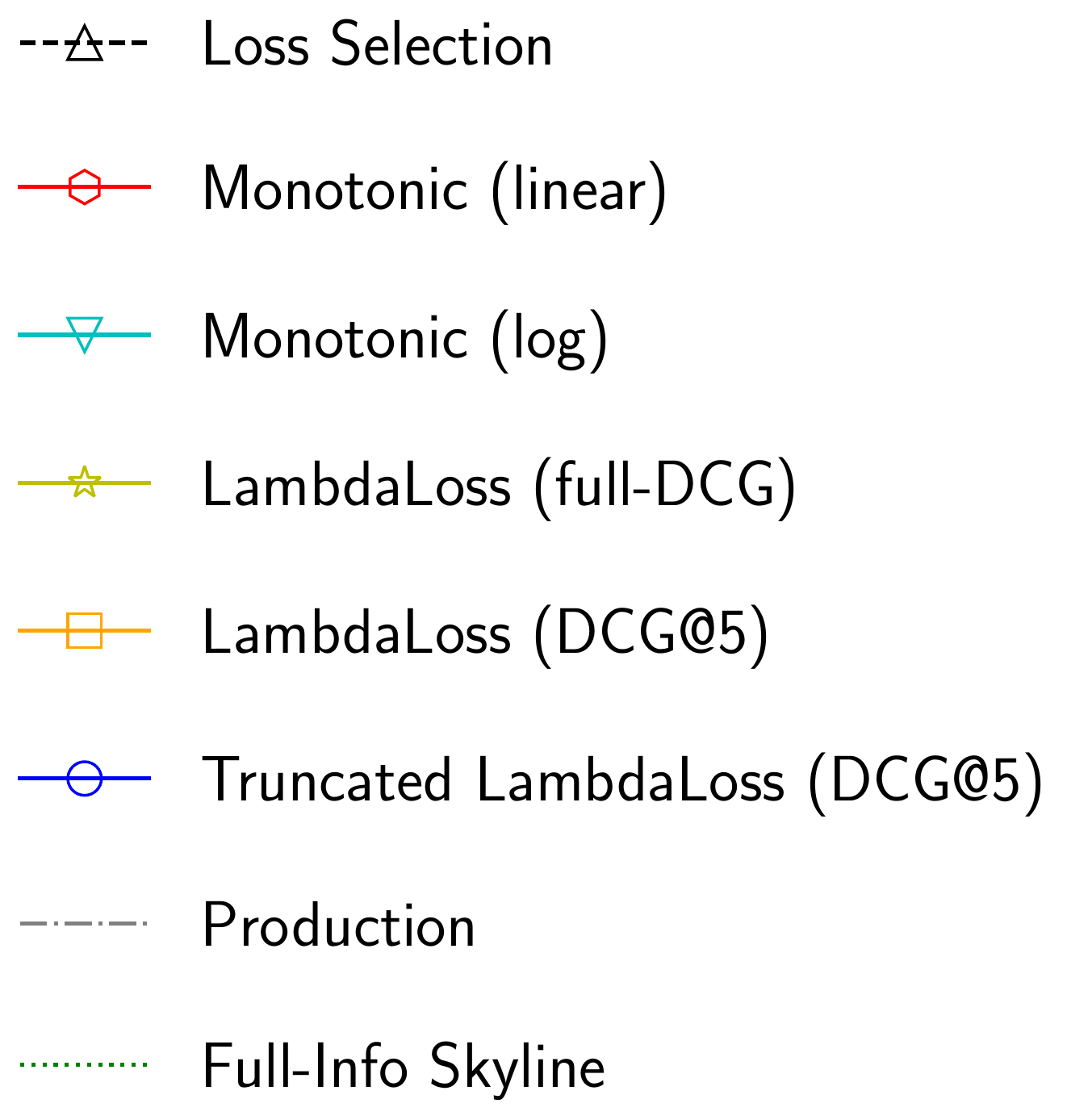}} 
\vspace{-0.3em}
\\
\includegraphics[scale=0.31]{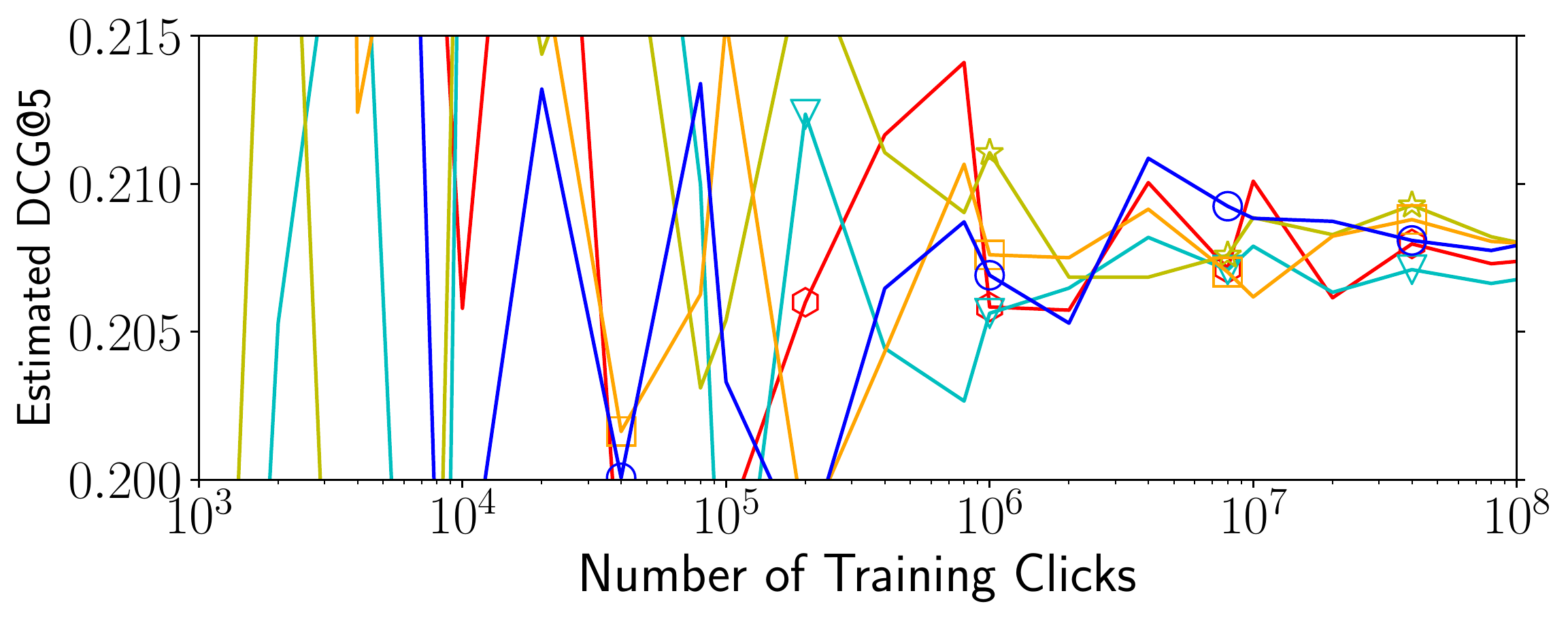} &
\includegraphics[scale=0.31]{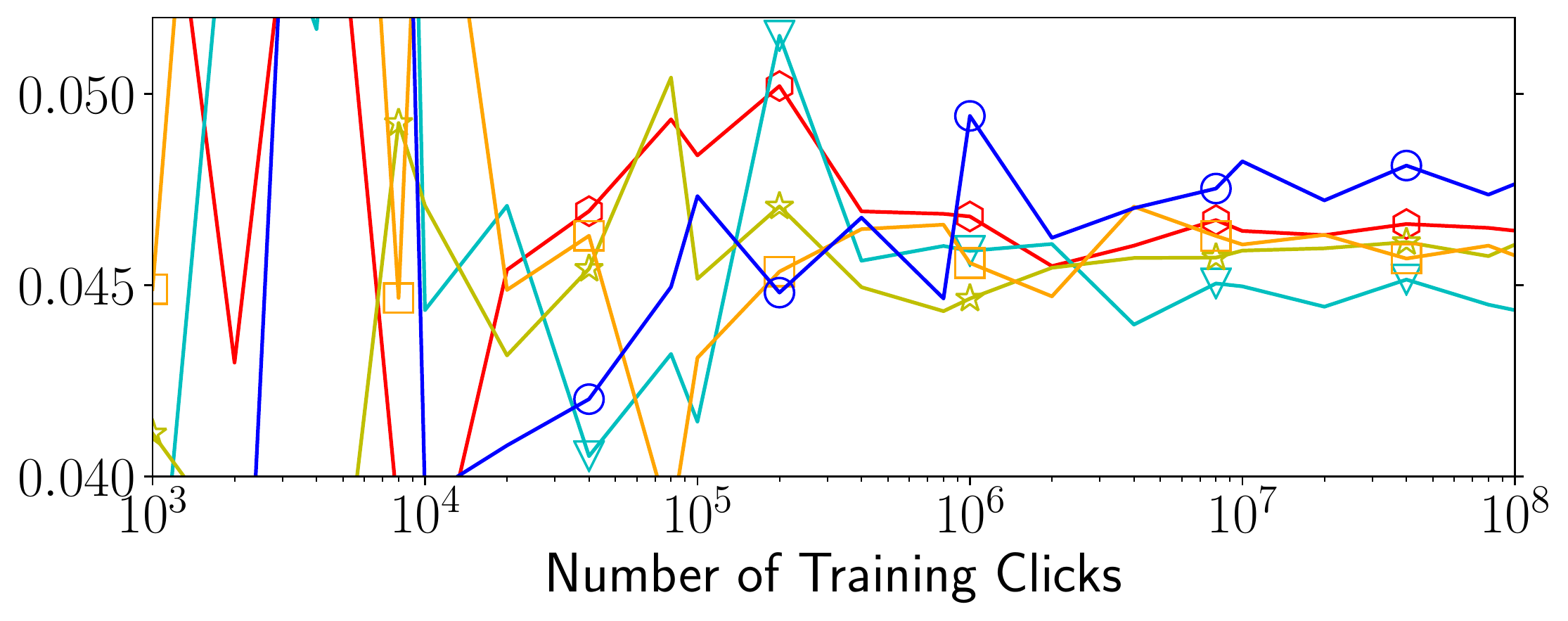} 
\end{tabular}
\caption{
Performance of the policy-aware estimator (Eq.~\ref{eq:policyaware}) optimizing \ac{DCG}$@5$ using different loss functions.
The loss selection method selects the estimated optimal model based on clicks gathered on separate validation queries.
Varying numbers of clicks on top-5 rankings with randomization, the number of validation clicks is 15\% of the number of training clicks.
}
\label{fig:dcglosses}
\end{figure*}

\subsection{Optimizing Top-$k$ Metrics}

Next, we consider the question: \emph{Are our novel counterfactual \ac{LTR} loss functions effective for top-$k$ \ac{LTR} metric optimization?}
Figure~\ref{fig:dcglosses} shows the performance of the policy-aware approach after optimizing different loss functions under top-$5$ feedback.
While on the Yahoo dataset little differences are observed, on the MSLR dataset substantial differences are found.
Interestingly, there seems to be no advantage in optimizing for \ac{DCG}$@5$ instead of full \ac{DCG} with the LambdaLoss.
Furthermore, the monotonic loss function works very well with a linear upper bound, yet poorly when using the log upper bound.
On both datasets the heuristic truncated LambdaLoss loss function provides the best performance, despite being the only method without a theoretical basis.
When few clicks are available, the differences change; e.g., the monotonic loss function with a log upper bound outperforms the other losses on the MSLR dataset when fewer than $10^5$ clicks are available.

Finally, we consider unbiased loss selection; Figure~\ref{fig:dcglosses} displays both the performance of the selected models and the estimated performance on which the selections are based.
For the most part the optimal models are selected, but variance does cause mistakes in selection when few clicks are available.
Thus, unbiased optimal loss selection seems effective as long as enough clicks are available.

In conclusion, we answer our second question positively: our results indicate that the truncated counterfactual LambdaLoss loss function is most effective at optimizing \ac{DCG}$@5$.
Using this loss, our counterfactual \ac{LTR} method reaches state-of-the-art performance comparable to supervised \ac{LTR} on both datasets.
Alternatively, our proposed unbiased loss selection method can choose optimally between models that are optimized by different loss functions.

\section{Related Work}

Section~\ref{section:supervisedLTR} has discussed supervised \ac{LTR} and Section~\ref{section:counterfactualLTR} has described the existing counterfactual \ac{LTR} framework; this section contrasts additional related work with our policy-aware approach.

Interestingly, some existing work in unbiased \ac{LTR} was performed in top-$k$ rankings settings \citep{agarwal2019addressing, agarwal2019estimating, wang2018position, wang2016learning}.
Our findings suggest that the results of that work are affected by item selection bias and that there is the potential for considerable improvements by applying the policy-aware method.

\citeauthor{carterette2018offline}~\citep{carterette2018offline} recognized that counterfactual evaluation cannot evaluate rankers that retrieve items that are unseen in the interaction logs, essentially due to a form of item selection bias.
 Their proposed solution is to gather new interactions on rankings where previously unseen items are randomly injected.
Accordingly, they adapt propensity scoring to account for the random injection strategy.
In retrospect, this approach can be seen as a specific instance of our policy-aware approach.
In contrast, we have focused on settings where item selection bias takes place systematically and propose that logs should be gathered by any policy that meets Condition~\ref{eq:awarecond2}.
Instead of expanding the logs to correct for missing items, our approach avoids systematic item selection bias altogether.

Other previous work has also used propensity scores based on a logging policy and examination probabilities.
\citet{Komiyama2015} and subsequently \citet{lagree2016multiple} use such propensities to find the optimal ranking for a single query by casting the ranking problem as a multiple-play bandit.
\citet{li2018offline} use similar propensities to counterfactually evaluate ranking policies where they estimate the number of clicks a ranking policy will receive.
Our policy-aware approach contrasts with these existing methods by providing an unbiased estimate of  \ac{LTR}-metric-based losses, and thus it can be used to optimize \ac{LTR} models similar to supervised \ac{LTR}.

Lastly, online \ac{LTR} methods where interactive processes learn from the user~\citep{yue2009interactively} also make use of stochastic ranking policies.
They correct for biases through randomization in rankings but do not use an explicit model of examination probabilities.
To contrast with counterfactual \ac{LTR}, while online \ac{LTR} methods appear to provide robust performance~\citep{jagerman2019comparison}, they are not proven to unbiasedly optimize \ac{LTR} metrics~\citep{oosterhuis2019optimizing, oosterhuis2018differentiable}.
Unlike counterfactual \ac{LTR}, they are not effective when applied to historical interaction logs~\citep{hofmann2013reusing}.

\section{Conclusion}

In this work, we have proposed a policy-aware estimator for \ac{LTR}, the first counterfactual method that is unbiased w.r.t. both position bias and item selection bias.
Our experimental results show that existing policy-oblivious approaches are greatly affected by item selection bias, even when only small amounts are present.
In contrast, the proposed policy-aware \ac{LTR} method can learn from top-$k$ feedback without being affected by the choice of $k$.
Furthermore, we proposed three counterfactual \ac{LTR} approaches for optimizing top-$k$ metrics: two theoretically proven lower bounds on \ac{DCG}$@k$ based on monotonic functions and the LambdaLoss framework, respectively, and another heuristic truncated loss.
Additionally, we introduced unbiased loss selection that can choose optimally between models optimized with different loss functions.
Together, our contributions provide a method for learning from top-$k$ feedback and for top-$k$ metrics. 
To the best of our knowledge, this is the first counterfactual \ac{LTR} method that is unbiased in top-$k$ ranking settings.
Arguably, this work also serves to further bridge the gap between supervised and counterfactual \ac{LTR} methods, as we have shown that state-of-the-art lambda-based supervised \ac{LTR} methods can be applied to the state-of-the-art counterfactual \ac{LTR} estimators.

Future work in supervised \ac{LTR} could verify whether potential novel supervised methods can be applied to counterfactual losses.
A limitation of the policy-aware \ac{LTR} approach is that the logging policy needs to be known; future work could investigate whether a policy estimated from logs also suffices~\citep{liu2018breaking, li2018offline}.
Finally, existing work on bias in recommendation~\citep{Schnabel2016} has not considered position bias, thus we anticipate further opportunities for counterfactual \ac{LTR} methods for top-$k$ recommendations.

\begin{acks}
We want to thank the anonymous reviewers for their feedback and Rolf Jagerman and Jin Huang for their helpful comments.
This research was partially supported by the Netherlands Organisation for Scientific Research (NWO) under project nr 612.001.551 and by the Innovation Center for AI (ICAI).
All content represents the opinion of the authors, which is not necessarily shared or endorsed by their respective employers and/or sponsors.
\end{acks}

\section*{Code and data}
To facilitate the reproducibility of the reported results, this work only made use of publicly available data and our experimental implementation is publicly available at \url{https://github.com/HarrieO/2020topkunbiasedltr}.

\bibliographystyle{ACM-Reference-Format}
\bibliography{references}

\end{document}